\documentclass[sigconf]{acmart}

\AtBeginDocument{%
  \providecommand\BibTeX{{%
    Bib\TeX}}}

\copyrightyear{2023} 
\acmYear{2023} 
\setcopyright{rightsretained} 
\acmConference[MICRO '23]{56th Annual IEEE/ACM International Symposium on Microarchitecture}{October 28-November 1, 2023}{Toronto, ON, Canada}
\acmBooktitle{56th Annual IEEE/ACM International Symposium on Microarchitecture (MICRO '23), October 28-November 1, 2023, Toronto, ON, Canada}
\acmDOI{10.1145/3613424.3614251}
\acmISBN{979-8-4007-0329-4/23/10}

\usepackage{physics}
\usepackage{algorithmic}
\usepackage{graphicx}
\usepackage{textcomp}
\usepackage{xcolor}
\usepackage{enumitem}
\usepackage{float}
\usepackage{multirow}
\usepackage{tikz}

\newcommand{\un}[1]{\underline{#1}}

\newcommand{\design}{ERASER}
\newcommand{\designext}{\design+M}
\newcommand*\circled[1]{\tikz[baseline=(char.base)]{
            \node[shape=circle,draw,inner sep=0.5pt, fill=white, text=black] (char) {#1};}}

\newcommand{\ignore}[1]{}
\newcommand{\parity}{\mathrm{parity}}
\newcommand{\data}{\mathrm{data}}

\renewcommand{\Pr}{\mathbb{P}}

\definecolor{oldaliceblue}{HTML}{070A52}
\definecolor{aliceblue}{rgb}{0.14, 0.01, 0.8}
\definecolor{OliveGreen}{HTML}{1A4D2E}
\definecolor{orange}{HTML}{1A4D2E}
\usepackage{xcolor}
\usepackage{tcolorbox}
\tcbset{width=1\columnwidth,boxrule=1pt,colback=aliceblue,arc=1pt,left=2pt,right=2pt,boxsep=0.5pt}

\newtcolorbox{hintbox}[2][]
{
  colframe = oldaliceblue!100,
  colback  = oldaliceblue!5,
  boxsep=2pt,
  width=\dimexpr\columnwidth\relax, 
  coltitle = oldaliceblue!20!black,
  title    = #2,
  #1,
}

\begin{document}

\begin{abstract}
    Quantum error correction (QEC) codes can tolerate hardware errors by encoding fault-tolerant logical qubits using redundant physical qubits and detecting errors using parity checks.
    \textit{Leakage errors} occur in quantum systems when a qubit leaves its computational basis and enters higher energy states. These errors severely limit the performance of QEC due to two reasons. \textit{First}, they lead to erroneous parity checks that obfuscate the accurate detection of errors. \textit{Second}, the leakage spreads to other qubits and creates a pathway for more errors over time. Prior works tolerate leakage errors by using leakage reduction circuits (LRCs) that modify the parity check circuitry of QEC codes. Unfortunately, naively using LRCs \textit{always} throughout a program is sub-optimal because LRCs incur additional two-qubit operations that (1)~facilitate leakage transport, and (2)~serve as new sources of errors. 

    Ideally, LRCs should \textit{only} be used if leakage occurs, so that errors from both leakage as well as additional LRC operations are simultaneously minimized. However, identifying leakage errors in real-time is challenging. To enable the robust and efficient usage of LRCs, we propose {\it \design\ } that speculates the subset of qubits that \textit{may} have leaked and only uses LRCs for those qubits. Our studies show that the majority of leakage errors typically impact the parity checks. We leverage this insight to identify the leaked qubits by analyzing the patterns in the failed parity checks. We propose ERASER+M that enhances \design\ by detecting leakage more accurately using qubit measurement protocols that can classify qubits into $\ket{0}, \ket{1}$ and $\ket{L}$ states. \design\ and \designext\ improve the logical error rate by up to $4.3\times$ and $23\times$ respectively compared to always using LRC.
\end{abstract}

\title{\design: Towards Adaptive Leakage Suppression \\for Fault-Tolerant Quantum Computing}

\author{Suhas Vittal}
\email{suhaskvittal@gatech.edu}
\orcid{0000-0003-0236-701X}
\affiliation{%
  \institution{Georgia Institute of Technology}
  \city{Atlanta}
  \state{GA}
  \country{USA}
}

\author{Poulami Das}
\email{poulami.das@utexas.edu}
\orcid{0000-0002-5811-6108}
\affiliation{%
  \institution{The University of Texas at Austin}
  \city{Austin}
  \state{Texas}
  \country{USA}
}

\author{Moinuddin Qureshi}
\email{moin@gatech.edu}
\orcid{0000-0002-1314-9096}
\affiliation{%
  \institution{Georgia Institute of Technology}
  \city{Atlanta}
  \state{GA}
  \country{USA}
}

\renewcommand{\shortauthors}{Vittal, Das, and Qureshi}
\acmArticleType{Research}
\acmCodeLink{https://github.com/borisveytsman/acmart}
\acmDataLink{htps://zenodo.org/link}

\maketitle

\section{Introduction}
Noisy quantum hardware and imperfect operations limit us from running most promising quantum applications~\cite{peruzzo2014variational, farhi2014quantum, reiher2017nitrogenfixation, shor1999polynomial, harrow2009hhl, childs2018firstqsim, campbell2019qalgsconstraintsat, kivlichan2020qsimelectrons, gidney2021rsafactorization, lee2021qchemtensorcontract, lemieux2021manybodyprep}. Quantum Error Correction (QEC) can bridge the gap between quantum applications and qubit devices. \textit{Fault-Tolerant Quantum Computers (FTQCs)} use QEC codes to encode \textit{logical qubits} using several \textit{physical qubits} such that the error rate of the logical qubits is lower than the physical error rate if the latter is below a certain \textit{threshold}. Moreover, the logical error rate decreases exponentially with increasing redundancy, measured as a QEC code's \textit{distance ($d$)}. This exponential suppression of errors enables QEC codes to achieve the target error rate necessary to run a given quantum application.  

In this paper, we focus on \textit{surface codes}, widely recognized as the most promising QEC code, which uses \textit{data qubits} to store quantum information and \textit{parity qubits} to detect errors~\cite{kitaev1997toriccodes, dennis2001surfacecodes, kitaev2003fault, fowler2012surface}. An FTQC executes \textit{syndrome extraction circuits} that project errors on the data qubits onto the parity qubits and measure the parity qubits to obtain a bitstring of parity checks called a \textit{syndrome}. A classical \textit{decoder} uses the syndrome to identify errors. It sends the correction to the \textit{control processor}, which corrects the errors. In practice, syndrome extraction uses quantum operations, which are also error-prone. To tolerate these errors, a decoder analyzes at least $d$ consecutive rounds of syndromes, also known as a \textit{QEC cycle}. FTQCs enable computations by interleaving QEC cycles in between logical operations.

\begin{figure*}
    \centering
    \includegraphics[width=\textwidth]{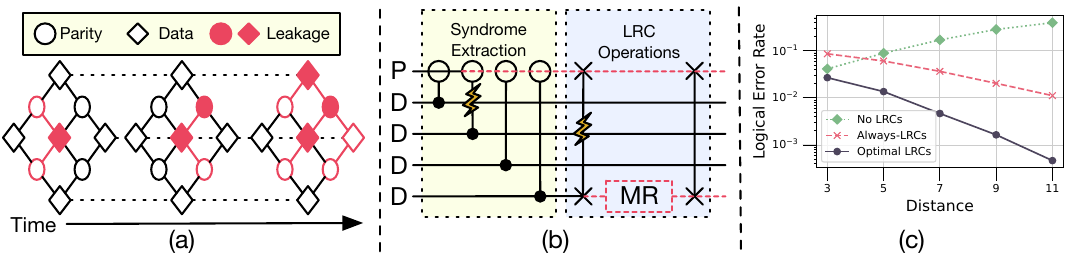}
    \caption{(a)~Leakage errors spread over time (b)~Regular syndrome extraction resets parity qubits every round, removing any leakage from them. Leakage reduction circuits (LRCs) swap data and parity qubits to remove leakage from the data qubit at the expense of five extra CNOTs (two extra SWAPs cost five CNOTs). (c) Logical error rate without LRCs, state-of-the-art Always-LRCs, and idealized LRC scheduling over 10 QEC cycles.}
    \label{fig:intro_figure}
\end{figure*}

Recent studies from IBM and Google show that \textit{leakage errors} degrade QEC performance on real hardware~\cite{ibm2022peekskillexperiments, google2023suppressing, krinner2022distance3demonstration}. 
Leakage errors occur when qubits leave the computational basis ($|0\rangle$ and $|1\rangle$) and enter a higher energy state $|L\rangle$~\cite{brown2019leakagemixedqubit, brown2019leakagesubsystem, brown2020leakagesurfacecode, ibm2022peekskillexperiments, google2023suppressing, suchara2015leakagesuppression, mcewen2021removingleakage, miao2022overcomingleakage, fowler2013leakagetopologicalcodes, manabe2023qutritleakagesims}. As quantum operations are only calibrated for the computational basis, leakage errors deteriorate logical performance for two reasons. \textit{First}, leaked qubits cause faulty operations during syndrome extraction, inducing random errors on their neighboring qubits and obfuscating other errors from being detected due to incorrect parity checks. 
\textit{Second}, these faulty operations spread leakage onto other qubits via \textit{leakage transport}. If not removed, leakage continues spreading, affecting more qubits over time and increasing the \textit{leakage population ratio}, or the number of qubits leaked at any given time, as shown in Figure~\ref{fig:intro_figure}(a), making QEC codes increasingly vulnerable. For example, our studies show that leakage errors increase the logical error rate by 27$\times$ and 467$\times$ for a distance 7 surface code after one and five QEC cycles, respectively. \textit{Thus, reducing the impact of leakage errors is crucial to improving the performance of QEC. }

Leakage errors arise from fundamental device-level imperfections and cannot be wholly eliminated despite improving qubit qualities. Instead, recent approaches actively remove them as they occur by resetting the leaked physical qubits. The most common technique uses leakage reduction circuits (LRCs) 
that modify the syndrome extraction circuit to \textit{swap} the data and parity qubits~\cite{aliferis2005lrcs, fowler2013leakagetopologicalcodes, suchara2015leakagesuppression, brown2020leakagesurfacecode}, as in Figure~\ref{fig:intro_figure}(b). Syndrome extraction rounds without LRCs proceed normally, where the parity qubits are measured and reset, eliminating any leakage from them. These rounds are followed by rounds with LRCs where the SWAPs remove leakage from the data qubits. Prior works schedule LRCs every alternate round throughout the duration of a program. \textit{However, our studies show that always scheduling LRCs is sub-optimal and limits their efficacy. }

\textit{Always-LRCs} scheduling throughout program execution done in prior work has the following drawbacks. \textit{First}, leaked qubits facilitate leakage transport through the two-qubit operations intended to eliminate them. Our studies show that the \textit{leakage population ratio (LPR)} continues to increase over QEC cycles despite using LRCs; ideally, we want the LPR to remain as low as possible to prevent performance degradation. \textit{Second}, LRC operations increase the number of two-qubit operations in a syndrome extraction round from $4$ to $9$, as shown in Figure~\ref{fig:intro_figure}(b). Two-qubit operations are themselves error-prone and serve as new sources of errors even when there are no leakage errors. Our studies show that although the state-of-the-art \textit{Always-LRCs} scheduling policy can improve the logical error rate, its performance is still far from an idealized policy that schedules LRCs only when leakage occurs. For example, Always-LRCs scheduling can improve the logical error rate by 4$\times$ for distance $7$ surface codes, as shown in Figure~\ref{fig:intro_figure}(c). However, the idealized policy can improve the logical error rate by up to 10$\times$. Furthermore, this gap consistently increases with the increasing code distance, heavily limiting the performance of QEC. This paper aims to bridge this gap via the optimal usage of LRCs such that leakage errors are maximally removed while limiting leakage spread and minimizing errors from LRC operations. We propose \textit{\design} that achieves this goal.

\design\ comprises of three key components. (1)~The \textit{Leakage Speculation Block (LSB)} analyzes the current syndrome to speculate potentially leaked qubits, (2)~the \textit{Dynamic LRC Insertion (DLI)} block modifies the next syndrome extraction round to include LRC operations for this subset of qubits, and (3)~the \textit{QEC Schedule Generator (QSG)} issues the updated syndrome extraction schedules to the qubits. Designing efficient LSB logic is non-trivial because leakage errors may remain invisible during syndrome extraction while continuing to induce errors on other qubits. Even if they impact syndrome extraction, they create random parity qubit flip patterns, as a leaked data qubit can cause any arbitrary combination of its four neighboring parity qubits to flip. Efficient DLI logic design is also challenging as the QEC schedules must be adapted in real time. Failure to introduce LRCs in real-time causes leakage to persist, whereas waiting to determine which LRCs to use causes QEC cycles to slow down and errors to accumulate on qubits. 

To overcome these challenges, we leverage the insight that most leakage errors become visible to syndrome extraction within a few syndrome extraction rounds, and thus, optimizing the LSB to tackle visible leakage is sufficient. To address the challenge related to arbitrary syndrome bit-flip patterns caused by leakage errors, we speculate a leakage has occurred if at least half of the neighboring parity qubits have flipped for a data qubit. This achieves a sweet spot between two extremes: a conservative prediction based on too few neighboring syndrome bit-flips introduces more LRC operations, whereas a more aggressive prediction may cause leakage to remain undetected. Note that during Always-LRCs scheduling, each data qubit swaps with a unique parity qubit. However, as \design\ schedules LRCs dynamically, two data qubits may request to swap with the same parity qubit, thus preventing their LRCs from being scheduled concurrently. To resolve this problem, DLI schedules the LRCs in the upcoming round to maximize the number of LRCs scheduled. By scheduling LRCs only for likely-leaked data qubits, \design\ removes leakage errors while minimizing any additional errors caused by LRCs.

\begin{figure*}
    \centering
    \includegraphics[width=\textwidth]{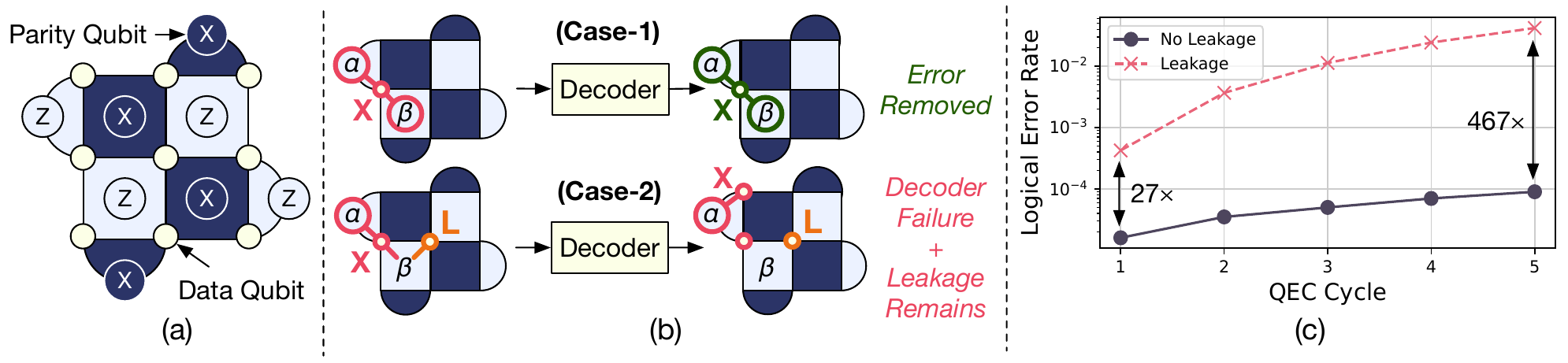}
    \caption{(a) Distance ($d$) 3 surface code.
    (b) In Case~1 (no leakage), there is an $X$ error on a data qubit that causes two $Z$ stabilizers ($\alpha$ and $\beta$) to flip. 
    In Case~2 (with leakage), the same data qubit has an $X$ error, but the central data qubit has a leakage error ($L$), which causes stabilizer $\beta$ not to flip. The decoder fails to identify the actual data qubit error and instead assigns the $X$ error to the boundary. (c) Logical performance comparison with and without leakage errors.
    }
    \label{fig:bg_figure}
\end{figure*}

Finally, recent device-level research has been increasingly exploring the efficacy of \textit{multi-level} readout, which classifies additional states beyond $\ket{0}$ and $\ket{1}$~\cite{chen2023transmon2st3stdisc, marques2023uwavelrctransmon}. While the accuracy of multi-level readout is worse than standard readout, the additional information granted by multi-level readout can enhance leakage detection accuracy. We propose \designext\ that leverages multi-level readout to improve LRC scheduling further.

Our evaluations show that \design\ and \designext\ improve the logical error rate by up to $4.3\times$ and $23\times$, respectively, compared to Always-LRCs scheduling. Furthermore, \design\ requires <1\% logic and 5ns latency on Xilinx FPGAs, demonstrating that real-time leakage suppression can be achieved at low cost. Further evaluations regarding the applicability of \design\ to alternatives for LRCs and an analysis of its performance under different noise models can be found in the Appendix.

\vspace{0.05in}
Overall, this paper makes the following contributions:

\begin{enumerate}[ leftmargin=0cm,itemindent=.5cm,labelwidth=\itemindent,labelsep=0cm,align=left, itemsep=0.2 cm, listparindent=0.5cm]
    \item Our studies show that always scheduling LRCs throughout program execution (Always-LRCs scheduling) has limited efficacy. 
    
    \item We propose {\it \design},\ a dynamic LRC scheduling policy that predicts the subset of qubits that may have leaked and only applies LRCs to those qubits in real time. To the best of our knowledge, this is the \textit{first} paper that proposes real-time leakage suppression.

    \item We propose a \textit{Leakage Speculation Block} to accurately detect leakage and \textit{Dynamic LRC Insertion} to adapt the QEC schedules. 

    \item We propose {\it \designext}, which extends \design\ to leverage the capabilities of multi-level readout.
\end{enumerate}

\section{Background and Motivation}

\subsection{The Surface Code}
    A distance $d$ surface code encodes a logical qubit using an alternating lattice of $d^2$ data and $d^2-1$ parity qubits, as shown in Figure~\ref{fig:bg_figure}(a). Periodically, \textit{syndrome extraction} circuits entangle the data qubits with their neighboring parity qubits to project any errors on the data qubits into \textit{Pauli} errors or $I$ (none), $X$ (bit flip), $Y$ (bit and phase flip), and $Z$ (phase flip) errors on the parity qubits~\cite{fowler2012surface, kitaev1997toriccodes, dennis2001surfacecodes, kitaev2003fault, nielsenchuang}. Each parity qubit and its neighboring data qubits execute a quantum circuit to measure a 4-qubit operator, called a \textit{stabilizer}, in a \textit{syndrome extraction round}. The surface code uses two types of stabilizers, $X$ and $Z$, which correct $Z$ and $X$ errors, respectively. The surface code can correct arbitrarily many errors provided the errors do not form an \textit{error chain} (a sequence of adjacent errors) of length more than $\lfloor (d-1) / 2 \rfloor$.

\subsection{Decoding Errors on the Surface Code}
    Errors on the logical qubit are detected by periodically executing syndrome extraction circuits, which measure the parity checks. These parity checks, also called \textit{syndromes}, are used to identify errors on the logical qubit in \textit{real-time} by pairing or matching the non-zero parity bits using graph algorithms, such as \textit{Minimum-Weight Perfect Matching (MWPM)}. This process is known as decoding and is performed independently for $X$ and $Z$ stabilizers. For example, matching the non-zero or flipped $Z$ syndrome bits $\alpha$ and $\beta$ enables the decoder to accurately identify the $X$ error on the data qubit shown in \textit{Case-1} of Figure~\ref{fig:bg_figure}(b). In practice, decoders simultaneously decode $d$ consecutive rounds of syndromes to tolerate operational errors in syndrome extraction. This constitutes a \textit{logical} or \textit{QEC cycle}. Amongst decoders for the surface code, MWPM is widely recognized as the gold standard for decoding surface codes because of its high accuracy~\cite{vittal2023astrea, das2022lilliput, das2022afs, ravi2022clique, higgott2021pymatching}.

\subsection{Pauli+ Errors and Leakage Errors}
    Not all errors on data qubits can be classified as Pauli errors on real quantum hardware. In recent demonstrations of QEC codes, Google identified a class of errors in addition to decoherence and operational errors that reduce the performance of QEC~\cite{google2023suppressing}. These errors are referred to as \textit{Pauli+} errors and are fundamentally hard to tackle for two reasons. First, it is difficult to characterize these errors in real systems. Second, these errors can cause correlated errors, and thus, a decoder unaware of such correlations may be unable to handle such errors~\cite{yu2014lookup, brown2020leakagesurfacecode, google2023suppressing}.

    \textit{Leakage errors}, which occur when a qubit leaves the computational basis ($\ket{0}$ and $\ket{1}$) and enters a higher energy state $\ket{L}$, are the most damaging class of Pauli+ errors because leaked physical qubits spread errors onto other qubits through two-qubit operations~\cite{brown2019leakagemixedqubit, brown2019leakagesubsystem, brown2020leakagesurfacecode, fowler2013leakagetopologicalcodes, aliferis2005lrcs, varbanov2020leakagedetection, bultink2020protectingleakage, miao2022overcomingleakage, google2023suppressing, manabe2023qutritleakagesims}. As these two-qubit operations are calibrated for only the computational basis, performing an operation between a leaked qubit and an unleaked qubit can lead to either (1)~a random error modifying the unleaked qubit's state (which can be modeled as a random Pauli error), or (2)~the unleaked qubit becoming leaked through \textit{leakage transport} from the leaked qubit~\cite{miao2022overcomingleakage, manabe2023qutritleakagesims, varbanov2020leakagedetection, marques2023uwavelrctransmon}.
    
    Although leakage errors are less frequent than gate and measurement errors, they significantly degrade the logical performance because the errors spread by leakage errors obfuscate other errors from getting detected. For example, \textit{Case-2} of Figure~\ref{fig:bg_figure}(b) shows how the leaked qubit at the center of the code lattice leads to faulty syndrome extraction on its adjacent $Z$ stabilizer $\beta$, causing it not to flip. Now, the decoder observes only a single non-zero $Z$ stabilizer $\alpha$ and assigns an $X$ error to the data qubit on the boundary of the lattice, thereby introducing an error itself while the actual $X$ error remains undetected. Moreover, the leaked qubit remains faulty, creating a pathway for the leakage to spread onto other qubits, leading to an even greater possibility of errors in future syndrome extraction rounds. Consequently, the logical error rate increases, degrading the performance of QEC.
    
    For example, Figure~\ref{fig:bg_figure}(c) shows the logical error rate of a distance $7$ surface code over multiple QEC cycles. After the first cycle, the logical error rate is 27$\times$ higher in the presence of leakage. Moreover, the impact of leakage accumulates over multiple cycles. 
    For example, the logical error rate increases by only $5\times$ after five QEC cycles in the absence of leakage errors, whereas it increases by nearly $100\times$ in the presence of leakage errors. Leakage errors rapidly widen the gap in logical performance with increasing QEC cycles, going from $27\times$ to $467\times$ in just five cycles. The sharp decline in logical performance shows that \textit{leakage errors pose a significant barrier to scaling up logical qubits and realizing fault tolerance.}

\subsection{Prior Works on Leakage Error Reduction} 
\begin{figure}[!htb]
        \centering
        \includegraphics[width=0.95\columnwidth]{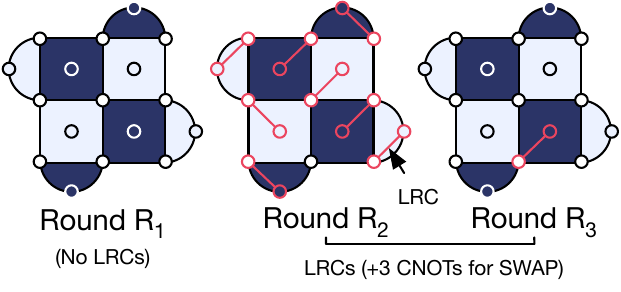}
        \caption{An example of a SWAP LRC schedule.}
        \vspace{-0.1in}
        \label{fig:swaplrc}
    \end{figure}

    There are several prior works that focus on leakage error mitigation that can be classified into three broad categories:

    \begin{enumerate}[ leftmargin=0cm,itemindent=.5cm,labelwidth=\itemindent,labelsep=0cm,align=left, itemsep=0.2 cm, listparindent=0.5cm]
     \item \textbf{Post-processing}: This approach identifies leakage errors from stabilizer flips observed during syndrome extraction~\cite{bultink2020protectingleakage, varbanov2020leakagedetection}. The drawback of this approach is that it requires many rounds to determine leakage errors accurately, and thus, it is mainly used to post-select or filter trials that had leakage errors during memory experiments on real systems.

    \item \textbf{Calibrating operations on leaked states}: This approach mitigates leakage by using new operations on leaked qubits that interact with states ($\ket{L}$) outside the computational basis~\cite{mcewen2021removingleakage, miao2022overcomingleakage, marques2023uwavelrctransmon}. Such approaches are either inherently specific to the underlying quantum processor~\cite{miao2022overcomingleakage} or require calibrating custom pulses to interact with higher energy states~\cite{marques2023uwavelrctransmon, lacroix2023fastfluxlru}. Thus, such approaches are not the focus of this paper, which tackles leakage in a manner \textit{generalizable} to any processor.\footnote{We include results for ERASER combined with such specialized operations in the Appendix (Section~\ref{sec:dqlrstudy}).}
    
    \item \textbf{SWAP-Based Leakage Removal}: This involves swapping \textit{leaked data} qubits with \textit{unleaked parity} qubits during syndrome extraction. The modified syndrome extraction circuit is called a \textit{leakage reduction circuit} or \textit{LRC}~\cite{suchara2015leakagesuppression, aliferis2005lrcs, brown2020leakagesurfacecode, fowler2013leakagetopologicalcodes}.\footnote{In this paper, LRCs refer to SWAP LRCs. While there are other variants of LRCs which we discuss in Related Work (Section~\ref{sec:relatedwork}), these are impractical because they require denser device connectivity.} The measurement and reset operations post-SWAP eliminate leakage from the data qubit. LRCs are scheduled periodically to minimize both parity and data qubit leakage, as shown in Figure~\ref{fig:swaplrc} for the $d=3$ code. In round $R_1$, no LRCs are performed, and the parity qubits are measured and reset during usual syndrome extraction, thus removing any leakage from the $d^2-1$ parity qubits. In round $R_2$, $d^2-1$ data qubits are scheduled for LRCs (each data qubit is swapped with a unique parity qubit). The LRC in round $R_3$ removes leakage from the remaining data qubit. LRCs are a straightforward approach for leakage reduction readily implementable on any device as their only overhead is modifying the syndrome extraction circuit.

    \end{enumerate}

\subsection{Limitations of LRCs}
    LRCs have two fundamental limitations. \textit{First}, LRCs are unoptimized for reducing the impact of leakage transport as it has only been observed recently on real systems~\cite{mcewen2021removingleakage, miao2022overcomingleakage}. \textit{Second}, LRCs are inefficiently scheduled: qubits do not have leakage errors every round, so using LRCs only adds additional points of failure during syndrome extraction.

\subsection{Goal}
    Ideally, we want greater accuracy while maintaining the simplicity of LRCs to mitigate leakage errors. Our proposed solution \design\ achieves this goal.
\section{Are Always-LRCs a Good Idea?}
We discuss the limitations of LRCs, specifically their poor performance in the presence of leakage transport.

\subsection{LRCs facilitate leakage transport}\label{sec:leakageanalysis}
\begin{figure}[!htb]
        \centering
        \includegraphics[width=\columnwidth]{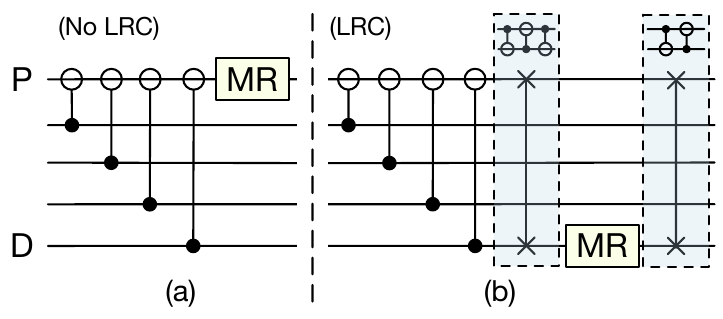}\vspace{-0.15in}
        \caption{Syndrome extraction (a)~without an LRC, and (b)~with an LRC. In (b), one SWAP swaps the parity and data qubit states, and another swaps them back.}
        \label{fig:swaplrccircuit}
    \end{figure}

    An LRC, shown in Figure~\ref{fig:swaplrccircuit}(b), removes leakage from a data qubit ($D$) by swapping it with a parity qubit ($P$). However, an LRC may introduce leakage onto $P$ via leakage transport when $D$ is leaked. In such a situation, the LRC may introduce leakage rather than remove it as intended. In the following section, we model the introduction of leakage errors during syndrome extraction with and without an LRC. A summary of the notation used in this section is shown in Table~\ref{tab:eqnotation}. 

    \begin{table}[!htb]
        \centering
        \begin{center}
        \caption{Notation and Constants Used in Section~\ref{sec:leakageanalysis}}
        \label{tab:eqnotation}
        \begin{tabular}{|c|c|}
            \hline
            Notation & Explanation \\
            \hline
            \hline
            \multirow{2}{*}{$\Pr(L_\data | L_\parity)$} & Probability that a data qubit leaks \\
            & given a parity qubit is already leaked. \\
            \hline
            \multirow{2}{*}{$\Pr(L_\parity | L_\data)$} & Probability that a parity qubit leaks \\
            & given a data qubit is already leaked. \\
            \hline
            \multirow{2}{*}{$p_\ell$} & Probability of CNOT leakage error, \\
            & equal to $0.1p = 1 \times 10^{-4}$. \\
            \hline
            \multirow{2}{*}{$p_\mathrm{LT}$} & Probability of CNOT leakage transport, \\
            & equal to 0.1. \\
            \hline
        \end{tabular}
        \end{center}
    \end{table}

    \subsubsection{Leakage Errors Without LRCs}
    Consider a syndrome extraction round without an LRC, as shown in Figure~\ref{fig:swaplrccircuit}(a), and suppose that the parity qubit $P$ is leaked. During the round, $P$ may transport leakage to one of its neighboring data qubits, which we denote $D$, whereas any leakage on $P$ will be removed once it is reset. Thus, we are interested in the probability that $D$ \textit{becomes leaked by the end of the round, given} $P$ \textit{is leaked before the start of the round}. We denote this probability as $\Pr(L_\data | L_\parity)$ as designated in Table~\ref{tab:eqnotation}.
    
    $D$ can only incur leakage through either (1)~operation errors through CNOTs with neighboring parity qubits or (2)~a transport error in the CNOT with $P$. For calculating (1), we note that the probability of the $k$-th CNOT causing leakage (not due to transport) is $(1-p_\ell)^{k-1}p_\ell$, and so (1) is the sum of these probabilities for $1 \leq k \leq 4$. Then, combining (1) and (2), $\Pr(L_\data | L_\parity)$ is found as in Equation~\eqref{eqn:p2d}, and we estimate this quantity to be about $10\%$.
    \begin{equation}
        \label{eqn:p2d}
        \Pr(L_\data | L_\parity) = p_\mathrm{LT} + \sum_{k=1}^4 (1-p_\ell)^{k-1}p_\ell
    \end{equation}

    \subsubsection{Leakage Errors with LRCs}
    Now, we consider syndrome extraction with an LRC, as in Figure~\ref{fig:swaplrccircuit}(b), and suppose that now the data qubit $D$ is leaked. We are interested in the probability $P$ becomes leaked by the end of the round, given $D$ is leaked before the start of the round. We denote this probability as $\Pr(L_\parity | L_\data)$ as in Table~\ref{tab:eqnotation}. However, unlike the situation without an LRC, we note that $P$ is used in nine CNOTs. Furthermore, $P$ interacts with $D$ six times. However, only \textit{four} of these CNOTs occur before $D$ is reset and can thus cause leakage transport. The other two CNOTs occur after $D$ is reset and are unlikely to cause leakage transport.
    
    As with the prior calculation, we can separate $\Pr(L_\parity | L_\data)$ into (1)~a probability of leakage caused by operation error and (2)~a probability of leakage caused by leakage transport. Thus, $\Pr(L_\parity | L_\data)$ is found as in Equation~\eqref{eqn:d2p}, which we estimated to be about $34\%$.
    \begin{equation}
        \Pr(L_\parity | L_\data) = \sum_{k=1}^9 (1-p_\ell)^{k-1}p_\ell + \sum_{k=1}^4 (1-p_\mathrm{LT})^{k-1}p_\mathrm{LT}
        \label{eqn:d2p}
    \end{equation}
    
    \subsubsection{Impact of Leakage Transport}
\begin{figure}[!h]
        \centering
        \includegraphics[width=\columnwidth]{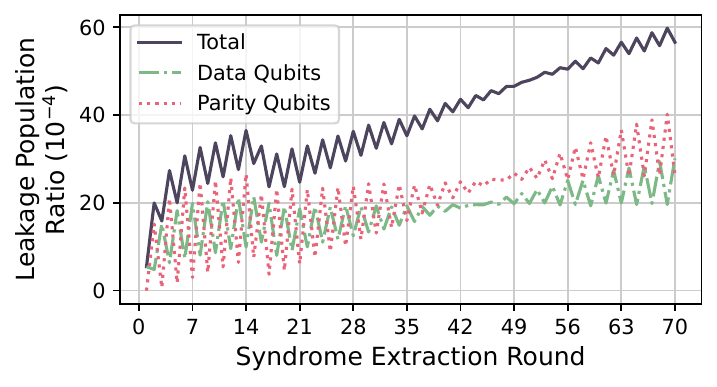}
        \caption{Leakage population ratios for $d=7$ at $p=1\times 10^{-3}$ over $70$ rounds ($10$ QEC cycles). Lower is better.}
        \label{fig:lrc_lpr}
    \end{figure}

    As $\Pr(L_\parity | L_\data)$ is about $3\times$ larger than $\Pr(L_\data | L_\parity)$, we expect that LRCs significantly contribute to increasing the amount of leakage on a logical qubit. This is indeed the case. Figure~\ref{fig:lrc_lpr} shows the \textit{leakage population ratio (LPR)}, or the probability that a given physical qubit on the logical qubit is leaked, over $10$ QEC cycles in a $d=7$ code at $p = 1 \times 10^{-3}$. We observe two trends corroborating our analytical results in Equation~\eqref{eqn:p2d} and Equation~\eqref{eqn:d2p}. \textit{First}, the LPR spikes after even rounds, which are rounds with LRCs. \textit{Second}, each spike generally increases the LPR over the last spike, increasing the LPR over time.

\subsection{LRCs are inefficiently scheduled} 
    The state-of-the-art LRC policy schedules LRCs every alternate round such that rounds without LRCs remove leakage from parity qubits, and rounds with LRCs remove leakage from data qubits. However, such scheduling is inefficient because the additional CNOTs in LRCs create new sources of failure. Ideally, we want to use LRCs only to remove leakage errors when they occur. 
    
\begin{figure}[!htb]
        \centering        
        \includegraphics[width=\columnwidth]{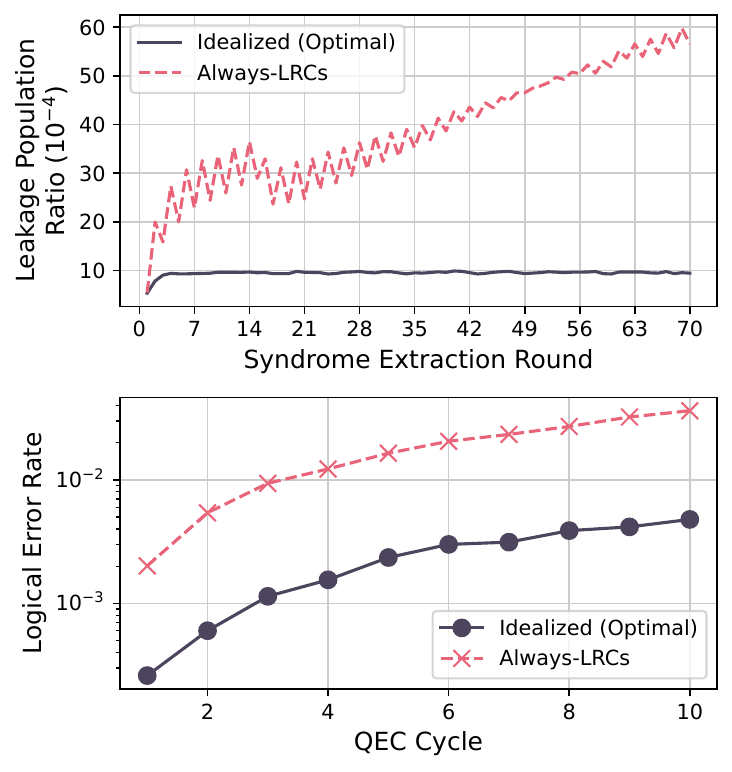}
        \vspace{-0.15in}
        \caption{LPR (top) and LER (bottom) comparison between state-of-the-art and idealized LRC scheduling.}
        \vspace{-0.15in}
        \label{fig:lrcvsideal}
    \end{figure}

    To assess the impact of the extra LRC operations on the logical performance (LPR and LER), we compare state-of-the-art LRC scheduling to an idealized scheduling policy that schedules LRCs for qubits as soon as they are leaked. Figure~\ref{fig:lrcvsideal} shows the LPR and LER for both policies over 10 QEC cycles for a $d=7$ code at $p=1 \times 10^{-3}$. The LPR continues to increase for the state-of-the-art policy, resulting in 10$\times$ higher LER than the idealized policy. The performance gap is due to the idealized policy scheduling significantly fewer LRCs: the idealized policy schedules \textit{one LRC every three QEC cycles} whereas the state-of-the-art policy schedules \textit{24 LRCs every round}.

\subsection{Characterizing the Spread of Leakage}
\begin{figure}[!htb]
    \centering
    \includegraphics[width=0.85\columnwidth]{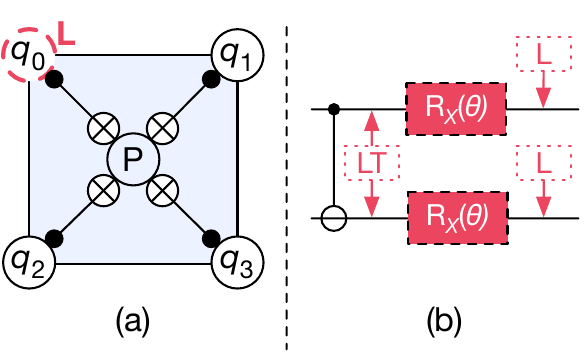}
    \color{blue}
    \caption{(a)~The simulated $Z$ stabilizer. The density matrix simulation starts with $q_0$ initialized in $|2\rangle$. (b)~CNOTs are followed by leakage transport, $R_X$ errors, and leakage injection.}
    \vspace{-0.05in}
    \label{fig:dmsimoverview}
\end{figure}

To better understand how leakage spreads on a real system, we perform density matrix simulations of a $Z$ stabilizer on the surface code. Our simulation implements the leakage phenomena observed by Google during their recent demonstration of a distance $5$ surface code on their Sycamore processor~\cite{google2023suppressing, miao2022overcomingleakage}. As Google Sycamore's leakage phenomena are reported to interact with the $|3\rangle$ state, our simulation uses \textit{ququarts},\footnote{Ququarts are a quantum superposition of $|0\rangle$, $|1\rangle$, $|2\rangle$, and $|3\rangle$.} where $|L\rangle$ corresponds to $|2\rangle$ and $|3\rangle$. Figure~\ref{fig:dmsimoverview}(a) provides an overview of our simulation, which simulates the spread of leakage originating from a single leaked data qubit $q_0$ across a $Z$ stabilizer over an LRC round followed by a no-LRC round. During syndrome extraction, each CNOT can incur errors due to (1)~leakage transport, (2)~$R_X$ error on unleaked qubits if one operand is leaked, and (3)~leakage injection, as shown in Figure~\ref{fig:dmsimoverview}(b). The $R_X$ errors in our experiment are fixed to be $R_X(0.65\pi)$, which was the error measured during leakage studies on Google Sycamore~\cite{miao2022overcomingleakage}.

\begin{figure}[!htb]
    \centering

    \vspace{-0.15in}
    \includegraphics[width=\columnwidth]{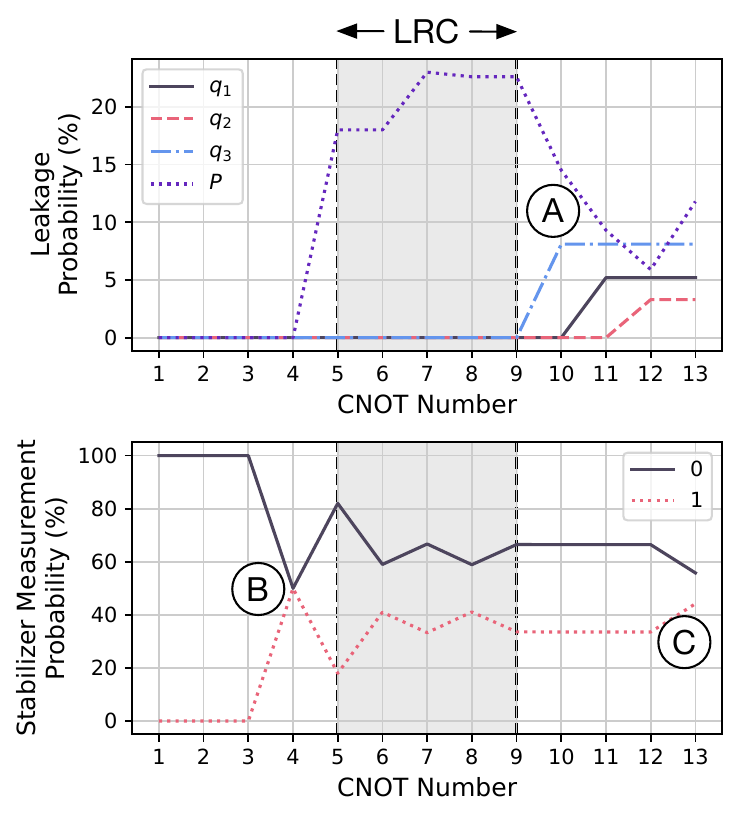}
    \color{blue}
    \vspace{-0.3in}
    \caption{(top)~Spread of leakage errors, and (bottom)~the effect of leakage on stabilizer measurement probability. We do not show qubit $q_0$'s leakage probability as it begins the simulation already initialized in $|2\rangle$.}
    \vspace{-0.05in}
    \label{fig:densitymatrixsim}
\end{figure}

Figure~\ref{fig:densitymatrixsim} shows the movement of leakage from the leaked parity qubit and the impact of leakage on the $Z$ stabilizer measurement. We discuss three points of interest. At point \circled{A}, which marks the end of the LRC with $q_0$, we observe that the parity qubit $P$ has significantly leaked due to interactions with $q_0$, confirming that \textit{LRCs do facilitate leakage transport}. Consequently, $P$ then spreads leakage errors onto the other data qubits during the no-LRC round, thus increasing the leakage population. Point \circled{B} shows the first point where $P$ is affected by leakage during a CNOT with $q_0$ (CNOT \#4). If $P$ was measured at this point, we would get a random outcome; note that we ideally want to measure $P$ as $0$ as there are no $X$ errors on the data qubits. As syndrome extraction continues, the measurement probabilities fluctuate. At point $\circled{C}$, before the measurement of $P$, the probability of measuring the correct outcome is slightly better than random. Thus, leakage errors interfere with syndrome extraction measurements by \textit{inducing random measurement results}.

We note that as our simulations in this section are restricted to a single stabilizer, the results observed \textit{understate} the impact of leakage. In reality, the leakage error on $q_0$ will also spread to the rest of the logical qubit and cause more errors. We refer to Google's recent studies on leakage and recently published qutrit simulations for more extensive analyses on the spread of leakage across an entire logical qubit~\cite{miao2022overcomingleakage, manabe2023qutritleakagesims}.

\section{\design: Insights and Design}

\begin{figure}[!htb]
    \centering
    \includegraphics[width=0.95\columnwidth]{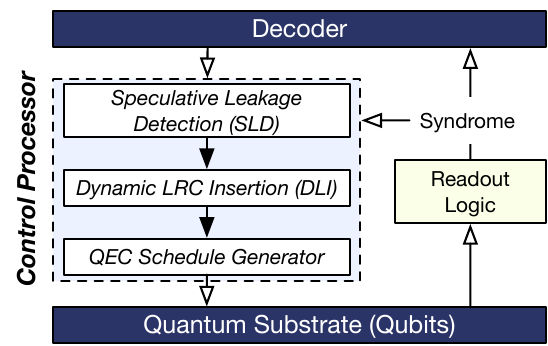}
    \caption{Overview of \design.}
    \label{fig:designoverview}
\end{figure}

We propose \design\ that judiciously schedules LRC operations such that errors from both leakage and LRC operations are simultaneously minimized. Figure~\ref{fig:designoverview} gives an overview of \design. The \textit{Leakage Speculation Block (LSB)} uses the current syndrome to speculate a subset of qubits that may have leaked. The \textit{Dynamic LRC Insertion (DLI)} block interrupts the \textit{QEC Schedule Generator (QSG)} to modify the syndrome extraction circuits for the next round and issues LRC operations only for qubits speculated as leaked.

Enabling adaptive LRCs presents two key challenges. \textit{First}, we must accurately speculate leakage. Failure to do so causes leakage to remain and degrade performance. \textit{Second}, the control processor must integrate LRC operations into QEC schedules in real time to prevent QEC cycles from stalling. We discuss the insights to overcome these challenges.

\begin{figure*}[!t]
    \centering
    \includegraphics[width=\textwidth]{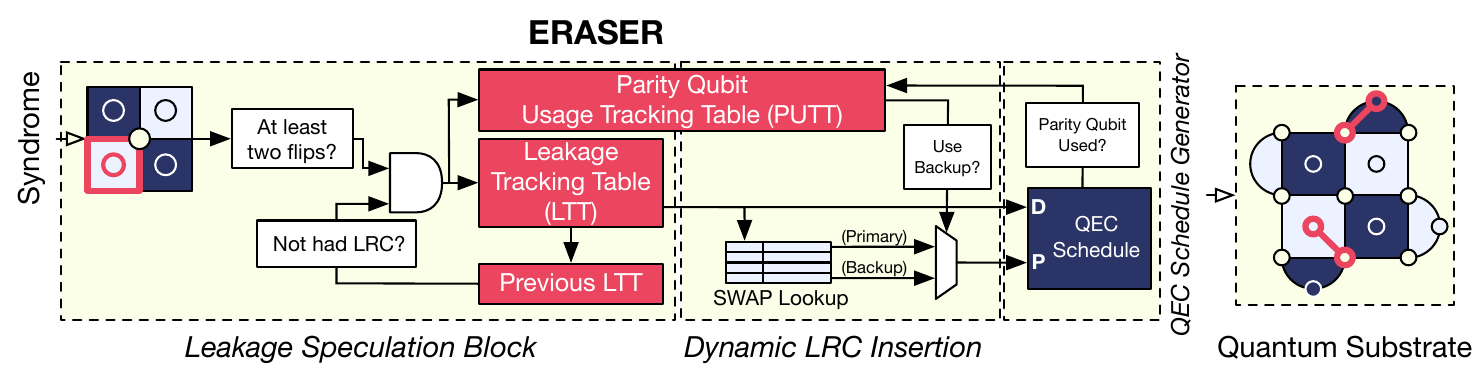}
    \caption{Overview of \design's microarchitecture.}
    \label{fig:design}
\end{figure*}
\subsection{Leakage Speculation Block: Challenges}
    The only information available about the qubits during syndrome extraction that could be used to detect leakage errors is the measured syndrome. We discuss how often leakage errors impact syndrome extraction and the challenges with precisely detecting leakage errors using syndromes.

    \subsubsection{Is Leakage Visible or Invisible from Syndromes?}
    Our analysis shows that leakage errors can be broadly classified into two categories: \textit{visible} which immediately affect syndrome extraction, and \textit{invisible} which persist for multiple rounds before affecting syndrome extraction. 
    We discuss how long a data qubit potentially remains invisible without LRCs; note that parity qubit leakage does not accumulate as these qubits are reset every round. This happens in two scenarios:
    \begin{enumerate}[ leftmargin=0cm,itemindent=.5cm,labelwidth=\itemindent,labelsep=0cm,align=left, itemsep=0.2 cm, listparindent=0.5cm]
     \item When the leaked data qubit causes an error in syndrome extraction. This can be modeled as a \textbf{depolarizing error} and affects parity qubit measurement with a 50\% probability.

    \item When the leaked data qubit \textbf{transports leakage} resulting in the parity qubit accumulating leakage. When measured, the parity qubit will be randomly classified as a 0 or 1. There is a 50\% probability that the error affects the measurement.
    \end{enumerate}

    As a data qubit neighbors at most four parity qubits, the probability a leaked data qubit is invisible in a round is $(\frac{1}{2})^4 = \frac{1}{16}$. As a qubit remains invisible until it affects a parity qubit measurement (probability is $1 - \frac{1}{16} = \frac{15}{16}$), the probability a leaked data qubit remains invisible for $r$ rounds is given by Equation~\eqref{eq:abc}.
    \begin{equation}
    \footnotesize
    \label{eq:abc}
        \Pr_\mathrm{invis}(r) = \frac{15}{16} \times \left( \frac{1}{16} \right)^r
    \end{equation}
    Table~\ref{tab:invisleakprob} shows the probability of a leaked data qubit remaining invisible over multiple rounds. Note that more than 99\% of leakage errors affect syndrome extraction within two rounds, resulting in most leakage errors becoming quickly visible.
    \begin{table}[!htb]
        \centering
        \begin{center}
        \vspace{-0.1in}
        \caption{Invisible Leakage Error Probability}
        \label{tab:invisleakprob}
        \begin{tabular}{|c|c|}
            \hline
            Rounds Spent Invisible & Probability ($\Pr_\mathrm{invis}$) in \%age \\
            \hline
            \hline
            0 & 93.8 \\
            \hline
            1 & 5.90 \\
            \hline
            2 & 0.36 \\
            \hline
            3 & 0.02 \\
            \hline
        \end{tabular}
        \vspace{-0.1in}
        \end{center}
    \end{table}
    
    \begin{hintbox}{\color{white}{\bf \design}: \textbf{Insight \#1}}
    Visible leakage errors are the \textit{most common variant} of leakage errors, and optimizing LRCs for them is sufficient.
    \end{hintbox}

\subsubsection{Challenges in Exact Leakage Detection}
    Syndrome bit flips not only result from leakage errors but also arise from other types of errors such as decoherence, gate, and measurement errors. This makes the reliance on syndromes to detect leakage errors extremely challenging. Furthermore, unlike other errors, leakage errors do not cause syndrome measurements to flip according to a specific pattern. For example, an $X$ error on a data qubit only causes its adjacent $Z$ syndromes to flip, whereas a measurement error causes the same syndrome measurement to flip across consecutive rounds. Unlike such errors, leakage errors cause random syndrome measurements to flip. For example, a leaked data qubit can cause any arbitrary combination of its four neighboring parity qubits to flip, making it difficult to detect the leakage during syndrome extraction.
    
    Instead of attempting to \textit{identify exactly where} leakage errors have occurred, we use the insight that leveraging the flipped syndrome bits to \textit{speculatively detect} a leakage with high accuracy is sufficient. However, even performing such speculative detection is nontrivial as there is an inherent trade-off between LRC scheduling frequency and performance. Speculating too conservatively schedules too many LRCs, degrading the QEC code's performance as the extra LRC operations increase errors during syndrome extraction. On the other hand, speculating too aggressively schedules LRCs too infrequently, also degrading performance as leakage is not removed in time. To maximize performance, \design\ achieves a sweet spot between the two and schedules LRCs on a data qubit when at least 50\% of its neighboring parity qubits flip. 

\begin{hintbox}{\color{white}{\bf \design}: \textbf{Insight \#2}}
    Speculative leakage detection has an inherent trade-off between LRC scheduling frequency and performance. Scheduling LRCs too aggressively or too conservatively will degrade performance by causing more errors to occur.
\end{hintbox}

\subsection{Leakage Speculation Block: Design}
    \design\ uses the current syndrome to speculate the data qubits that may have encountered leakage. The \textit{Leakage Speculation Block (LSB)} maintains a \textit{Leakage Tracking Table (LTT)} with one entry per data qubit, as shown in Figure~\ref{fig:design}. The LSB analyzes the current syndrome and speculates if a qubit has leaked. If it identifies a leakage, the corresponding LTT entry is marked as \textit{leaked}.\footnote{We use the term \textit{previous} round to indicate the last syndrome extraction round. We refer to the syndrome obtained from this round as the \textit{current} syndrome. The decision to introduce LRCs is made for the \textit{next} syndrome extraction round.} The LSB also maintains a \textit{Parity qubit Usage Tracking Table (PUTT)} to track the allocation of parity qubits for LRC operations on the data qubits. In the Always-LRCs scheduling, LRCs span two consecutive syndrome rounds as the number of data qubits exceeds the number of available parity qubits for swapping, and more than one data qubit may require swapping with the same parity qubit. As \design\ dynamically schedules LRC operations, the conflict is resolved by scheduling LRCs for the data qubits on any adjacent available parity qubit. The rules for handling the LTT and PUTT entries are discussed in the following subsection. The \textit{Dynamic LRC Insertion (DLI)} block uses the LTT and PUTT entries to introduce LRCs.

    \subsubsection{Speculating Leakage on Data Qubits}
        A data qubit may have two, three, or four neighboring parity qubits. If no LRC operations were scheduled for a data qubit in the previous round (which yields the current syndrome) \textit{and} at least half of the neighboring parity qubits flip, then the LSB block marks the LTT entry for the corresponding data qubit as \textit{leaked} to schedule LRC operations in the next round.  We choose half the number of parity qubits as a cutoff because, on average, half the parity qubits are expected to flip if there is a leakage error. Note that if LRC operations were scheduled on that particular data qubit in the previous round, any leakage on the qubit would have been removed, and we do not speculate any leakage even if 50\% of its neighboring parity qubits flip. 

    \subsubsection{Handling Parity Qubits Usage Tracking}
\begin{figure}[!htb]
            \centering
            \includegraphics[width=0.8\columnwidth]{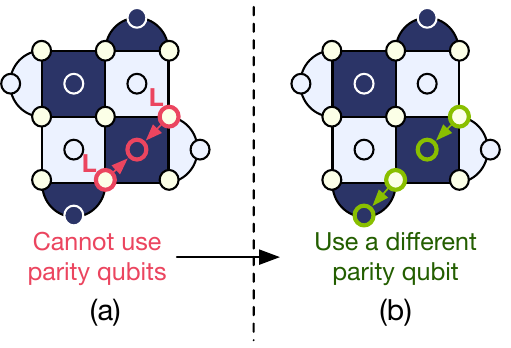}
            \caption{(a) Two leaked data qubits must perform an LRC but have the same primary parity qubit. (b) Arbitrarily assign data qubits to parity qubits.}
            \label{fig:resourcecontention}
        \end{figure}

        In Always-LRCs, each data qubit has a \textit{primary} parity qubit it swaps with to perform an LRC. However, as there are $d^2$ data qubits but only $d^2-1$ parity qubits, LRC operations cannot be scheduled for all data qubits in the same round. Instead, one LRC must be carried over into the next round. For example, Figure~\ref{fig:resourcecontention}(a) shows how both the leaked qubits conflict with the same primary parity qubit, and the LRC operations for both of them cannot be scheduled in the same round. To overcome this limitation, we leverage the insight that as \design\ schedules LRCs dynamically, only a subset of data qubits will require LRC operations in the same round. Thus, LRCs need not be carried over to the next round. To facilitate this, we select one of the neighboring parity qubits for LRC operations based on availability at runtime instead of allocating \textit{primary} parity qubits offline. The LSB allows each data qubit to use any neighboring parity qubit and marks it as \textit{used} in the PUTT. Now, the LRC operations for both leaked data qubits in Figure~\ref{fig:resourcecontention}(b) can be scheduled simultaneously.

        However, a completely arbitrary selection of parity qubits may lead to the accumulation of leakage on parity qubits if the same parity qubit gets selected for LRCs over multiple consecutive rounds for different data qubits. This happens because the associated parity qubit continuously gets swapped and is not reset for a prolonged duration. Note that each parity qubit may be used by up to four data qubits in case of such an arbitrary selection. To resolve this bottleneck, if a parity qubit has participated in an LRC in the previous round, it is marked as used in the PUTT and is not used for LRCs in the next round. The parity qubits that participated in LRC operations in the previous round will now be measured and reset in the next round, eliminating any leakage.
        The limited arbitrary selection of parity qubits enables us to schedule more LRCs in the same round and reduce leakage errors on both data and parity qubits. 

\subsection{Dynamic LRC Insertion: Challenges}
\begin{figure}[!htb]
        \centering
        \includegraphics[width=0.85\columnwidth]{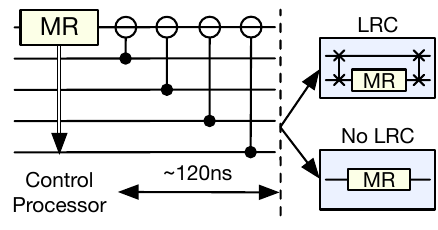}
        \caption{After a qubit is measured and the syndrome bit is sent to the control processor, there is a 120ns window to determine whether to schedule an LRC or not.}
        \label{fig:schedulingconstraint}
    \end{figure}

    Always-LRCs scheduling occurs offline before program execution by compiling syndrome extraction circuits down to the native gates of the quantum device~\cite{das2023imitationgame, mcewen2023timedynamics, smith2020ossqcompiler}. During program execution, the control processor repeatedly executes these gates in each syndrome extraction round. However, as \design\ only schedules LRCs when needed, it must interrupt the instruction supplier or the \textit{QEC Schedule Generator (QSG)} to update the schedule for the subset of qubits it has identified as leaked in the subsequent syndrome round. Note that the real-time constraint for scheduling ranges in the order of a few tens of nanoseconds. The QSG must know by the \textit{fourth} CNOT in the syndrome extraction circuit whether to schedule an LRC, as it will need to perform a SWAP after this CNOT to execute the LRC. Figure~\ref{fig:schedulingconstraint} shows this leaves about $120$ns between obtaining the previous syndrome and the end of the fourth CNOT in the current round, assuming each CNOT takes $30$ns (according to Sycamore latencies)~\cite{sycamoredatasheet, google2023suppressing}. Failure to resolve whether or not to introduce LRC operations within this time either causes the qubits to idle until a decision is made or moves the LRC operations to the next round, causing the leakage to remain. 
    Finally, as \design\ must be co-located within the control processor, it must fit on FPGAs to enable integration with existing quantum systems~\cite{sycamoredatasheet, ibmeagle, ibm2022sherbrooke, ibm2022quantumupdate, maurya2022nndiscreadout, das2022lilliput}.

\subsection{Dynamic Leakage Insertion: Design}
    After marking qubits as leaked in the LTT, \design\ attempts to schedule LRCs for all leaked data qubits while not scheduling parity qubits marked as used in the PUTT. We note that such scheduling is nontrivial as it requires solving a \textit{maximum matching} problem in real time. We must pair each leaked data qubit with a unique unused parity qubit to swap with during an LRC. Also, we must maximize the number of leaked data qubits scheduled for LRC operations to ensure all leaked data qubits are reset in the next round.
        
    To solve this problem efficiently, we propose using a lookup table containing pre-determined \textit{primary} and \textit{backup} SWAP neighbors for each data qubit; we call this lookup table the \textit{SWAP Lookup Table}. For each leaked data qubit, \design\ uses the SWAP Lookup Table to get a neighboring parity qubit to swap with. If the parity qubit is already marked as used in the PUTT, \design\ looks through the backup parity qubits and repeats this process. By default, our design maintains one backup parity qubit for each data qubit.

\subsection{QEC Schedule Generation: Design}
    After identifying LRCs that need to be scheduled, the control processor must execute the LRC operations in the next syndrome extraction round. By default, the control processor executes standard $X$ and $Z$ stabilizer circuits. The DLI interrupts the  QEC Schedule Generation (QSG), appends the instruction schedules by inserting the extra CNOTs corresponding to the LRC operations, and replaces the measurement operations on the associated parity qubits with those on the data qubits selected for LRC operations.

\subsection{Enhancing \design\ Using Multi-Level\\Readout Discriminators: \designext}
\begin{figure}[htp]
    \centering
    \includegraphics[width=\columnwidth]{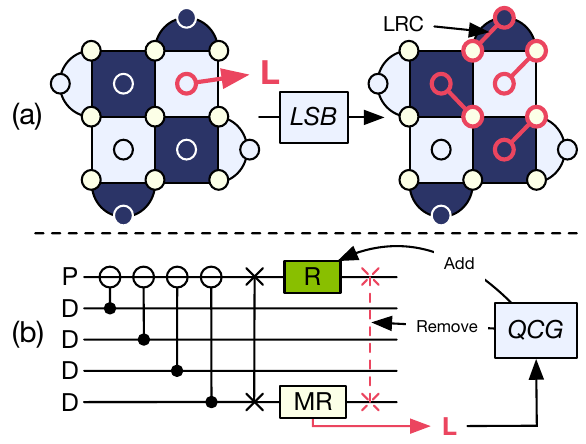}
    \caption{Key modifications for \designext. (a)~The LSB schedules LRCs for data qubits adjacent to a parity qubit measured in $\ket{L}$. (b)~The QCG modifies LRC operations upon measuring data qubit leakage.}
    \label{fig:mldoverview}
\end{figure}

The performance of \design\ is limited by the LSB's ability to detect leakage using syndromes. \design\ uses a passive leakage detection strategy because existing measurement discriminators are two-level classifiers that can only classify a qubit into states $\ket{0}$ and $\ket{1}$. Consequently, leakage is never actively detected because a leaked qubit is randomly classified into a 0 or 1. Multi-level discriminators, on the other hand, can also classify leaked states, and recent studies at the device level indicate they are promising for tackling leakage~\cite{krinner2022distance3demonstration, chen2023transmon2st3stdisc}. We propose \designext\ to integrate multi-level discriminators with \design\ for enabling more accurate leakage detection and LRC scheduling. It requires modifications to the LSB and QSG blocks, as are discussed next and summarized in Figure~\ref{fig:mldoverview}(a, b), respectively. Note that the DLI block does not require any modifications. 

\subsubsection{Modifications to the LSB}
If a parity qubit is classified as $\ket{L}$ in the current round, we assume it has transported leakage to one or more of its neighboring data qubits. Therefore, we speculate all its adjacent data qubits have been potentially leaked and mark the corresponding entries in the LTT so that LRC operations can be scheduled on these qubits in the next round.

\subsubsection{Modifications to the QSG}
During syndrome extraction with an LRC, if the data qubit is classified as $\ket{L}$, we observe that the parity qubit has a meaningless state since the SWAP during the LRC would have failed due to the data qubit leakage. Either the parity qubit has leaked 
or has a random unleaked state. Consequently, performing the SWAP after the data qubit reset is unnecessary as no useful information will be returned to the data qubit. However, the SWAP was also the only way to return the parity qubit to $\ket{0}$, and this must be done before the next round. Thus, if the data qubit is classified as $\ket{L}$, the QSG (1)~schedules a reset operation on the parity qubit and (2)~squashes the second SWAP in the LRC circuit.

\section{Evaluation Methodology}

In this section, we discuss our evaluation methodology before discussing our results.

\subsection{Surface Code Parameters}
We consider \textit{rotated surface codes} with code distances ($d$) ranging from $d=3$ to $d=11$. Rotated codes have lower resource overheads ($2d^2 - 1$ qubits) compared to the \textit{unrotated codes} ($(2d-1)^2$)~\cite{yu2014lookup, horsman2012latticesurgery} and, therefore, have been used in recent QEC studies and real-system demonstrations~\cite{google2023suppressing, miao2022overcomingleakage}.

\subsection{Error Model}
In this subsection, we discuss the error model used in our evaluations corresponding to different types of errors. 
\subsubsection{Modeling Operation Errors}
    We consider a physical error rate of $p = 1 \times 10^{-3}$ and a \textit{circuit-level} error model that injects (1)~\textit{depolarizing} errors on data qubits with probability $p$ at the start of a round, (2)~\textit{measurement} errors on qubits with probability $p$, (3)~\textit{depolarizing} errors on qubit operands after each CNOT or $H$ gate with probability $p$, and (4)~\textit{initialization} errors on qubits after a reset with probability $p$~\cite{argonneQCtutorial, gidney2021stim}.
    
\subsubsection{Modeling Leakage Errors}\label{sec:leakageerrormodel}
    Modeling leakage in memory experiments is inherently challenging to approximate in a \textit{tractable} manner~\cite{suchara2015leakagesuppression, google2023suppressing, manabe2023qutritleakagesims}. To ensure our results are reflective of real systems, we design our leakage error model based on prior studies on real systems~\cite{google2023suppressing, varbanov2020leakagedetection, miao2022overcomingleakage, ibm2022peekskillexperiments, marques2023uwavelrctransmon}. Our simulations also inject and track leakage in a manner consistent with prior work~\cite{fowler2013leakagetopologicalcodes, suchara2015leakagesuppression, brown2020leakagesurfacecode, brown2019leakagesubsystem, manabe2023qutritleakagesims}.
    
    We extend the circuit-level error model to inject leakage errors (1)~on data qubits at the beginning of each round with probability $0.1p$ to model \textit{\textbf{environment-induced}} leakage and (2)~on qubit operands after CNOT operations with probability $0.1p$ to model \textit{\textbf{operation-induced leakage}}.\footnote{Our error model also implements seepage, or the return of a leaked qubit to the computational basis, at the same rate as leakage. If a qubit is leaked, it can return to the computational basis in a randomly initialized state with probability $0.1p$.} When an unleaked qubit interacts with a leaked qubit through a CNOT, we inject a random Pauli error ($I, X, Y, Z$) on the unleaked qubit and apply a \textit{\textbf{leakage transport}} with a $10\%$ probability. 
    
    Our implementation of leakage transport conservatively assumes that the source qubit remains leaked after a leakage transport. Section~\ref{sec:altltmodel} reports results for an alternative implementation of leakage transport, where the source qubit may return to the computational basis provided the other qubit is not leaked.
    
\subsubsection{Modeling the Measurement of Leaked Qubits}
    The output state of a qubit (0 or 1) is determined by a measurement discriminator~\cite{jurvevic2021qv64, maurya2022nndiscreadout, ibm2022peekskillexperiments}. 
    If a standard two-level discriminator measures a leaked qubit, the outcome will be random because the discriminator is not trained to classify $|L\rangle$. We assume this for \design. For \designext, we assume that a multi-level discriminator, which classifies $\ket{0}$, $\ket{1}$, and $\ket{L}$, is erroneous at a rate of $10p$ to be consistent with results on real systems~\cite{chen2023transmon2st3stdisc, marques2023uwavelrctransmon}.

\subsection{Simulation Infrastructure}
\ignore{ 
    We use Google's \textit{Stim} simulator~\cite{gidney2021stim}, a state-of-the-art framework for performing state-preservation, or \textit{memory}, experiments~\cite{byun2022xqsim, gidney2021honeycombcode, google2023suppressing, wu2022surfstitch, berent2022qldpcsoftware, gidney2022benchmarkinghoneycomb}, which we have extended to simulate leakage errors. A memory experiment involves (1)~initializing the logical qubit to $|0\rangle$, (2)~performing $d$ syndrome extraction rounds to obtain $d$ syndromes, and (3)~measuring the logical qubit. Then, the decoder uses the $d$ syndromes to identify a correction for the logical qubit. If the correction is consistent with the logical measurement, then the memory experiment is a success. Otherwise, there is a logical error. 
    
    We evaluate up to ten \textit{QEC cycles} (each cycle is $d$ rounds) to better evaluate the efficacy of our design across multiple QEC cycles. We use Minimum Weight Perfect Matching decoding~\cite{edmonds1965blossom, kolmogorov2009blossomv}, but any other decoder may be used as well.
}

   We use Google's \textit{Stim} simulator~\cite{gidney2021stim}, a state-of-the-art framework for performing state-preservation, or \textit{memory}, experiments~\cite{byun2022xqsim, gidney2021honeycombcode, google2023suppressing, wu2022surfstitch, berent2022qldpcsoftware, gidney2022benchmarkinghoneycomb}, which we have extended to simulate leakage errors. Our evaluations go up to ten \textit{QEC cycles} (each cycle is $d$ rounds) to evaluate the efficacy of our design over time. We use Minimum-Weight Perfect Matching decoding~\cite{edmonds1965blossom}, but any other decoder may be used as well.

\subsection{Evaluation Metrics}
We use the (1)~\textit{logical error rate} and (2)~\textit{leakage population ratio} to evaluate our policies. The logical error rate (LER) quantifies the ability to suppress errors~\cite{das2022afs, das2022lilliput, holmes2020nisqplus, ueno2021qecool, ueno2022qulatis, liao2023witgreedy, ueno2022neoqec, google2023suppressing, miao2022overcomingleakage, brown2020leakagesurfacecode, suchara2015leakagesuppression, ibm2022peekskillexperiments}. The LER is defined in Equation~\eqref{eq:ler}:
\begin{equation}
\label{eq:ler}
    LER = \frac{n_\mathrm{logical\_errors}}{n_\mathrm{experiments}}
\end{equation} 

Leakage population ratio (LPR) quantifies the number of leaked qubits at any time~\cite{brown2019leakagemixedqubit, brown2019leakagesubsystem, miao2022overcomingleakage, mcewen2021removingleakage}. This metric is widely used in the devices community and is defined in Equation~\eqref{eq:lpr}:
\begin{align}
\label{eq:lpr}
    LPR &= \frac{1}{n_\mathrm{experiments}} \times \sum_{\mathrm{experiments}} \frac{n_\mathrm{leaked}}{n_\mathrm{qubits}}
\end{align}

\subsection{Hardware Cost of \design\ }
To evaluate the hardware overheads of our design, we target Xilinx's off-the-shelf Kintex UltraScale+ FPGA and synthesize our design using Vivado.

\ignore{ 
\vspace{0.1in}
   \begin{hintbox}{\color{white}{\bf \design}: \textbf{Figure-of-Merit}}
    For performance, a lower LER and LPR are better and desirable. For practicality and feasibility of implementation, \design\ must fit on off-the-shelf FPGAs and dynamically schedule LRC operations in real-time (in less than 120 nanoseconds).
    \end{hintbox}
}


\section{Evaluations}
In this section, we discuss the performance of our proposed designs \design\ and \designext. 

\subsection{Impact on Logical Error Rate}
\begin{figure}[!htb]
    \centering
    \includegraphics[width=\columnwidth]{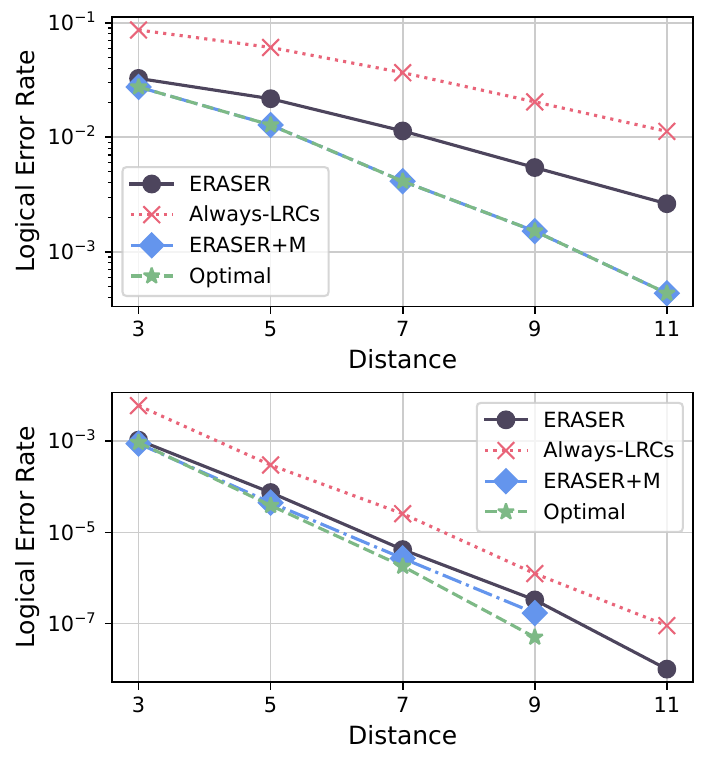}
    \caption{LER with increasing code distance for (top) $p = 10^{-3}$ and (bottom) $p = 10^{-4}$ for 10 QEC cycles. Data is not shown for $d = 11, p = 10^{-4}$ for \designext\ and optimal LRC scheduling as it was too low to be measured accurately.}
    \label{fig:lervsd}
\end{figure}

Logical error rate (LER) denotes the capability of a QEC code to suppress errors. A lower LER and exponentially decreasing LER with increasing code distance are desirable. Figure~\ref{fig:lervsd} shows \design\ improves the LER consistently with increasing distance on average by $3.3\times$ and up to $4.3\times$ in the best-case. \designext\ is even more effective and achieves near-optimal LER, improving the LER on average by $8.6\times$ and up to $26\times$ in the best case.

At lower physical error rates such as $p = 10^{-4}$, \design's performance improves, reducing the LER by $5.4\times$ on average and up to $9\times$ compared to Always-LRCs. Concurrently, \design's performance is now closer in performance to \designext\ and optimal scheduling, as error events become sparser at lower physical error rates~\cite{delfosse2022ufdecoder, delfosse2020hierarchical, ufd, smith2022cellularautomatapredecoder, chamberland2022hierarchical} and so leakage errors become more visible.

\ignore{
Figure~\ref{fig:ler_d11} further shows how the logical error rate changes over time for all policies for a $d = 11$ code at $p = 10^{-3}$. We observe that \design\ consistently maintains a lower LER by up to $6.3\times$. \designext\ bridges the gap between optimal LRC scheduling and almost matches the performance of the idealized setting, reducing the LER by up to $23\times$ compared to Always-LRCs and $5.5\times$ compared to \design. 

\begin{figure}[!htb]
    \centering
    \includegraphics[width=\columnwidth]{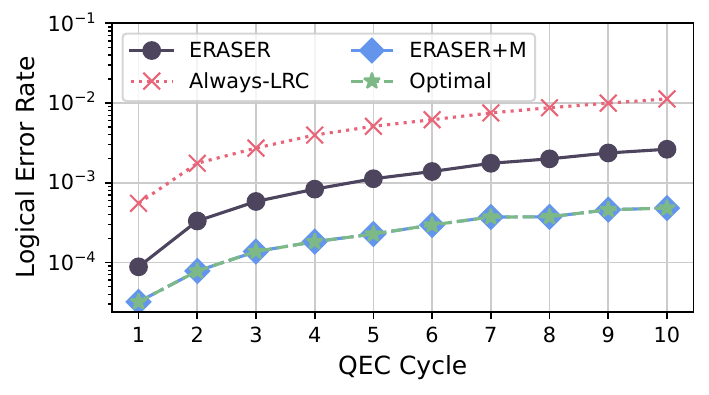}
    \caption{LER for different LRC scheduling policies assuming the default $d=11$ configuration.}
    \label{fig:ler_d11}
\end{figure}
}
\subsection{Impact on Leakage Population Ratio}
\begin{figure}[!htb]
    \centering
    \includegraphics[width=\columnwidth]{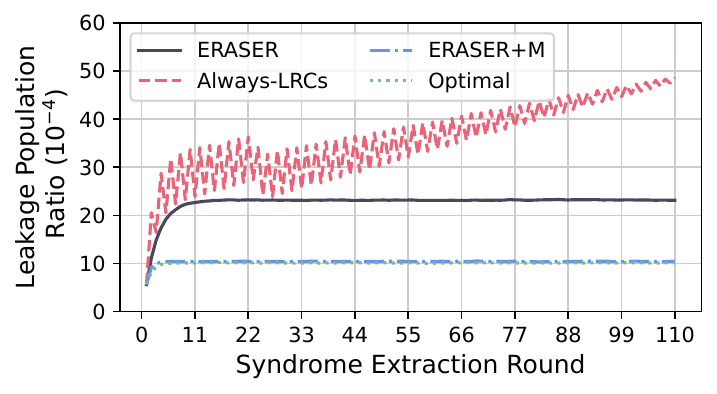}
    \caption{LPR for the baseline Always-LRCs, \design, \designext, and optimal (idealized) LRC scheduling for default $d=11$ configuration.}
    \label{fig:lpr_d11}
\end{figure}

A lower leakage population ratio or LPR means a greater reduction of leakage errors. Figure~\ref{fig:lpr_d11} shows the LPR of the default $d=11$ configuration for the competing LRC scheduling policies. \design\ consistently maintains a lower LPR and decreases the LPR by  $1.5\times$ on average and up to $2.1\times$. Furthermore, \designext\ bridges the gap between optimal LRC scheduling and reduces the LPR by $2.2\times$ compared to \design, performing nearly identically to the idealized setting of scheduling LRCs. 

\subsection{Hardware Implementation Cost}
Table~\ref{tab:hardware} shows the hardware resources required for implementing our proposed \design\ on standard off-the-shelf FPGAs as they are already being used to control and readout qubits on most existing quantum computers. Our implementation of \design\ requires less than 1\% logic utilization up to $d=11$ and has a worst-case latency of 5ns to speculate leakage and adapt the QEC schedules. This makes \design\ a very practical, low overhead, and accurate solution that eliminates leakage errors in real-time. 

\begin{table}[!htb]
    \centering
    \begin{center}
    \caption{FPGA Synthesis Results}
    \label{tab:hardware}
    \begin{tabular}{|c||c|c|}
        \hline
        $d$ & LUT (\%) & FF (\%) \\
        \hline
        \hline
        3 & 0.04 & 0.02 \\
        \hline
        5 & 0.12 & 0.05\\
        \hline
        7 & 0.26 & 0.10 \\
        \hline
        9 & 0.42 & 0.18 \\ 
        \hline
        11 & 0.76 & 0.26 \\
        \hline
    \end{tabular}
    \end{center}
\end{table}

\subsection{Performance Analysis of \design }
\begin{figure}[!htb]
    \centering
    \includegraphics[width=\columnwidth]
    {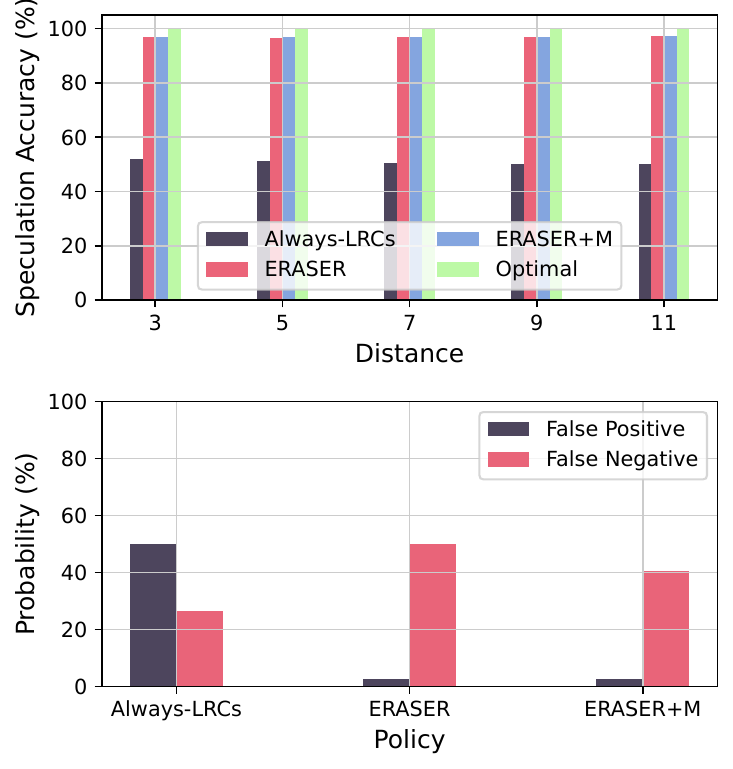}
    \vspace{-0.3in}
    \caption{(top) LRC speculation accuracy, and (bottom) FPRs and FNRs for $d=11$ over 10 QEC cycles. Data is not shown for optimal LRC scheduling as it has 100\% speculation accuracy.}
    \vspace{-0.1in}
    \label{fig:lrchitrate}
\end{figure}

\design\ is effective due to two key reasons. First, the LSB can accurately speculate most of the leakage errors. Second, \design\ schedules a significantly lower number of LRC operations. Figure~\ref{fig:lrchitrate}(a) shows the average speculation accuracy of the LSB. Both \design\ and \designext\ correctly use LRCs about 97\% of the time, whereas Always-LRCs correctly speculates about 50\% of the time. Table~\ref{tab:lrccost} further shows the average number of LRCs used per syndrome extraction round for all four policies. Both \design\ and \designext\ reduce the number of LRCs scheduled by $16.0\times$ on average and by up to $17.4\times$ in the best-case.

\begin{table}[!htb]
    \centering
    \begin{center}
    \caption{Average LRCs Used Per Round}
    \label{tab:lrccost}
    \begin{tabular}{|c||c|c|c|c|}
        \hline
        $d$ & Always-LRCs & \design & \designext & Optimal \\
        \hline
        \hline
        3 & 4.2 & 0.27 & 0.26 & 0.005\\
        \hline
        5 & 12 & 0.81 & 0.79 & 0.015 \\
        \hline
        7 & 24 & 1.52 & 1.50 & 0.034 \\
        \hline
        9 & 40 & 2.40 & 2.38 & 0.058\\ 
        \hline
        11 & 60 & 3.45 & 3.41 & 0.089\\
        \hline
    \end{tabular}
    \end{center}
\end{table}

\subsubsection{Examining the 3\% Gap} 
We further analyzed why there is a 3\% accuracy gap between \design\ (and \designext) and optimal LRC scheduling. Figure~\ref{fig:lrchitrate} shows the \textit{false positive rates (FPR)} and \textit{false negative rates (FNR)} for LRC usage across all policies. We make two observations. \design\ and \designext\ can easily identify situations with no leakage errors, with a 3\% FPR compared to a 50\% FPR for Always-LRCs. Minimizing FPR is crucial as qubits are typically not leaked, so applying LRCs may create new errors. However, \design\ is not as accurate when detecting leakage, though \designext\ can improve detection accuracy by up to 1.2$\times$.

\subsubsection{When does \design\ have False Negatives?}
The higher FNR of \design\ may appear alarming, but we observe that the false negatives incurred by ERASER are \textit{hard-to-detect} leakage errors. By design, \design's false negatives are either (1)~invisible leakage errors or (2)~leakage errors that only flip one parity check, which go undetected as \design\ schedules LRCs when at least \textit{two} parity checks have flipped. Such errors are hard to identify as they barely affect any syndrome measurements. Nevertheless, as shown with \designext, which has an FNR of 40\% compared to \design's 50\%, even small reductions in the FNR can significantly improve the logical error rate, as \designext\ has similar performance to optimal LRC scheduling.

\subsection{Analysis of Trade-Offs for \designext }
Although \designext\ is significantly more effective compared to \design, it incurs overheads of using multi-level discriminators. The measurement discriminator of a qubit is prepared by initializing it into each possible state that we want to classify, measuring it, and using the output signal to train a classification function. Typically, each execution is repeated for a few thousand trials. Multi-level discriminators must be trained to classify $\ket{L}$ states in addition to the usual $\ket{0}$ and $\ket{1}$. This results in two sources of overheads: (1)~we must calibrate a single-qubit operation that can initialize a qubit in a higher energy state (such as $\ket{2}$) and (2)~additional executions to prepare and measure a qubit in the higher energy state to obtain the output signal for the leaked state. This process is required for each qubit. Assuming calibrating a single-qubit operation takes about 1K shots and another 1K shots are required to calibrate the classifier for the $\ket{L}$ state, we require $2NK$ extra trials where $N$ is the number of physical qubits on the machine. Nevertheless, we note that other strategies also leverage multi-level discrimination, and thus \designext\ naturally synergizes with such strategies~\cite{suchara2015leakagesuppression}.

Note that \design\ is already very effective, and integrating the modifications needed for \designext\ can be managed solely in software. Hence, the choice of using \design\ versus \designext\ can be left to the programmer.

\section{Related Work}
\label{sec:relatedwork}
In this section, we discuss related work and compare or contrast as appropriate. 

\subsection{Leakage errors and their impact on QEC}
Improving device qualities and increasing system sizes have accelerated the demonstration of QEC codes in recent years~\cite{ibm2022peekskillexperiments, google2023suppressing, krinner2022distance3demonstration}. These real system studies reveal that \textit{leakage errors} significantly degrade the performance of QEC. For example, the studies performed on Google Sycamore rely on post-processing the results to eliminate experimental results from rounds with leakage errors. While post-processing can be used during experimentation, it cannot be used during program execution on a fault-tolerant quantum computer, where errors, including leakage errors, must be suppressed in real time. In contrast, \design\ actively removes leakage errors by efficiently scheduling leakage reduction circuits.

\subsection{Handling leakage errors}

Although strategies for mitigating leakage errors have been studied in the past, they are either low-cost but inaccurate or accurate with added overheads~\cite{krinner2022distance3demonstration, miao2022overcomingleakage, mcewen2021removingleakage, google2023suppressing, ibm2022peekskillexperiments}. \textit{Leakage Reduction Circuits (LRCs)} remove leakage from data qubits by executing SWAPs with other ancilla or parity qubits~\cite{aliferis2005lrcs, brown2020leakagesurfacecode, suchara2015leakagesuppression}. There are three varieties: \textit{Full LRCs}, \textit{Partial LRCs}, and \textit{SWAP LRCs}. As the former two variants of LRCs require denser device connectivity, we consider SWAP LRCs in this paper, which remove leakage errors from data qubits by swapping them with parity qubits.

Recent works have provided new leakage reduction strategies through custom operations that interact with states outside the computational basis~\cite{miao2022overcomingleakage, marques2023uwavelrctransmon}. While such operations may require modifications to the quantum system~\cite{marques2023uwavelrctransmon}, additional calibration overheads~\cite{lacroix2023fastfluxlru}, or are specific to the underlying device~\cite{miao2022overcomingleakage}, their performance is rather promising as they offer better performance than SWAP-based LRCs. Nevertheless, as such operations can also be erroneous and introduce leakage themselves, we observe that \design\ can improve the fidelity of such approaches as well, which we discuss at length in Section~\ref{sec:dqlrstudy}.

\section{Conclusion and Discussion}

Leakage errors present a significant barrier to realizing fault-tolerant quantum computing as they degrade the performance of quantum error correction (QEC) codes. These errors cause qubits to leave computational basis states and enter higher energy states. Leakage errors are not device-specific and have been observed in both superconducting processors~\cite{miao2022overcomingleakage, marques2023uwavelrctransmon, mcewen2021removingleakage, krinner2022distance3demonstration} and ion traps~\cite{brown2019leakagemixedqubit}.

Prior works actively eliminate these errors by using leakage reduction circuits (LRCs) to periodically remove leakage from data qubits through SWAPs and resets. However, always using LRCs throughout a program is sub-optimal as they introduce additional two-qubit operations that facilitate leakage transport onto other qubits and may themselves fail. Ideally, LRCs should be scheduled so that leakage is wholly removed while ensuring minimal impact from the extra LRC operations. 

We propose \design\ that detects the subset of qubits that may have leaked in real-time and judiciously applies LRC operations only on those qubits. \design\ leverages the insight that most leakage errors cause arbitrary parity check failures during QEC cycles. By identifying patterns in the failed parity checks, \design\ speculates the subset of leaked qubits. Once, the potentially leaked qubits are identified, \design\ adjusts the syndrome extraction schedules for these qubits by introducing LRC operations in real-time. The accuracy of leakage identification can be further enhanced by modifying the qubit measurement protocols to classify leaked states in addition to computational basis states. We leverage this insight to enhance \design\ using multi-level measurement classifiers. \design\ improves logical error rate by up to $4.3\times$ compared to Always-LRCs.

\design\ is the \textit{first} work to consider \textit{real-time leakage suppression}, and \design's superior performance to Always-LRCs demonstrates that real-time, or adaptive, leakage suppression provides significant benefits over static leakage suppression, where LRCs are scheduled offline at compile-time. Our results suggest that accurately speculating leakage in real-time is an important open problem. While \design's speculation accuracy is rather high, its poor FNR due to \textit{hard-to-detect} leakage errors is a significant source of logical error. Fortunately, we observe that even minor improvements in speculation accuracy, particularly in the FNR, can significantly improve the logical error rate. Thus, more sophisticated speculation strategies for leakage detection appear to be a rich and promising area for future research.

Finally, we observe that qubit loss in ion traps and neutral atom systems have a similar signature to leakage on superconducting systems, which was the predominant focus of this work~\cite{stricker2020iontraplossdetection, maksymov2022detectingfaultsoniontraps, wu2022erasurerydbergatomsnaqc, graham2022entonnaqc, covey2023qkdwithnaqcnodes}. As qubit losses can cause operations to fail, such systems must be capable of detecting qubit loss through loss detection mechanisms to avoid errors. Given this parallel, we expect strategies similar to \design\ may be fruitful on such systems. Furthermore, as ion traps and neutral atom systems are much slower than superconducting systems, we note that time constraints for identifying leaked qubits are more generous, allowing for more sophisticated and accurate strategies.

\begin{acks}
    We thank the reviewers of MICRO 2023 for their feedback. We thank Cody Jones (Google) for helpful discussions regarding leakage errors and for providing feedback on the submitted draft. We also thank Andrew Cross (IBM) and Adam Meier (GTRI) for providing feedback on the submitted draft. This research was conducted using the {\em Partnership for an Advanced Computing Environment (PACE)} cluster at Georgia Tech.
\end{acks}

\appendix
\section{Appendix: Additional Results}

\subsection{Alternative Model for Leakage Transport}\label{sec:altltmodel}
    The leakage transport model used in the main text conservatively assumes that the source qubit remains leaked after a transport; that is, both qubits involved in the transport are leaked after the transport finishes. In this section, we consider an alternative model, where the source qubit and receiving qubit ``exchange" leakage with each other. In such a model, if the receiving qubit is not leaked, it will become leaked, whereas the source qubit will return to the computational basis in a randomly initialized state. If the receiving qubit is leaked, then the transport essentially has no effect.

\begin{figure}[!htb]
        \centering
        \includegraphics[width=\columnwidth]{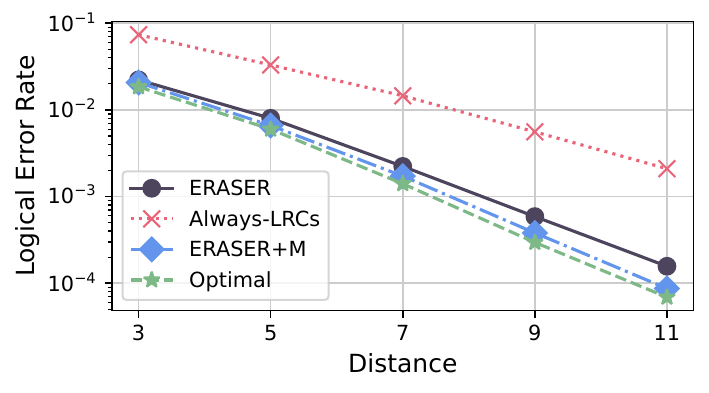}
        \caption{LER for $10$ QEC cycles at $p = 10^{-3}$ for the alternative leakage transport model.}
        \label{fig:altltmodeller}
    \end{figure}

    Figure~\ref{fig:altltmodeller} shows the LER for all policies under this alternative model for $10$ QEC cycles. As expected, all models improve quite a bit under the alternative model. However, we further note that the gap between \design\ and Always-LRCs \textit{widens} significantly, whereas the gap between \design\ and Optimal-LRCs has narrowed considerably. Now, \design\ improves the LER compared to Always-LRCs on average by $6.5\times$ and up to $13.4\times$. Concurrently, \designext\ improves the LER on average by $8.8\times$ and up to $24.1\times$.

\begin{figure}[!htb]
        \centering
        \includegraphics[width=\columnwidth]{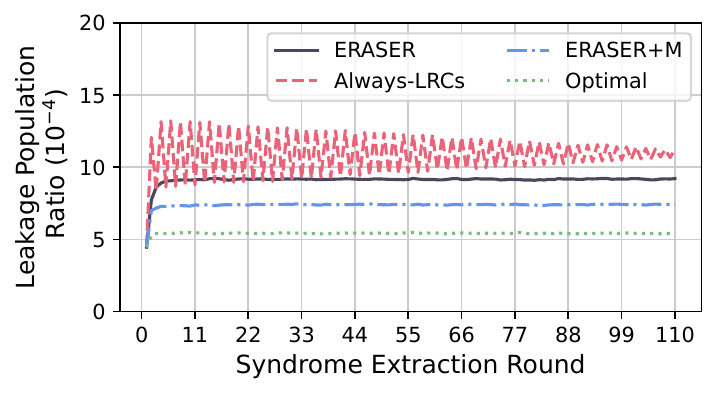}
        \caption{LPR over $110$ rounds for a $d = 11$ at $p = 10^{-3}$ using the alternative leakage transport model.}
        \vspace{-0.15in}
        \label{fig:altltmodellpr}
    \end{figure}

    We believe that \design\ significantly improves under this alternative leakage transport model for two reasons. \textit{First}, we note that the LPR for all policies is significantly lower. Figure~\ref{fig:altltmodellpr} shows the LPR for all four policies. We note that the LPR is substantially lower under the alternative model as the number of leaked qubits is preserved during a leakage transport under this model. Hence, the LPR curves for all policies, except Always-LRCs, stabilizes. The LPR for Always-LRCs spikes after rounds with LRCs and reduces after rounds without LRCs because LRCs may fail to remove leakage due to leakage transport. \textit{Second}, LRCs have lower error than in the original model. Consequently, the impact of \design's high FNR is much lower compared to the original model. 

\subsection{Applicability of \design\ with DQLR}\label{sec:dqlrstudy}
\begin{figure}
        \centering
        \includegraphics[width=\columnwidth]{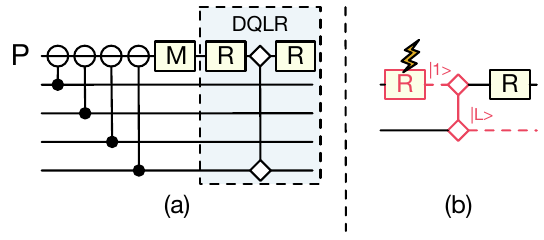}
        \caption{(a)~The DQLR protocol, which leverages the two-qubit LeakageISWAP operation. (b)~A reset failure on the parity qubit can cause the LeakageISWAP operation to excite the data qubit to $\ket{L}$.}
        \label{fig:dqlrcircuit}
    \end{figure}

    The evaluations in the main text consider the traditional SWAP-based LRC, which had been considered by much prior work~\cite{aliferis2005lrcs, fowler2012surface, suchara2015leakagesuppression, brown2020leakagesurfacecode}. However, recent work has been moving towards LRCs using custom operations with tremendous success~\cite{miao2022overcomingleakage, marques2023uwavelrctransmon}. As these operations exploit the underlying physics of the corresponding quantum processor, they can be calibrated for their respective systems without much difficulty. However, like any other quantum operation, these customized operations also may be erroneous. In this section, we analyze the applicability of \design\ for LRCs involving such operations. Specifically, we examine Google's \textit{DQLR} approach~\cite{miao2022overcomingleakage} as shown in Figure~\ref{fig:dqlrcircuit}(a), and we use the alternative leakage transport model from Section~\ref{sec:altltmodel} to ensure our results are reflective of Google Sycamore's leakage transport phenomena.

    \subsubsection{The DQLR Protocol}
        The DQLR protocol removes leakage from data and parity qubits every round by (1)~performing syndrome extraction as usual, (2)~resetting all parity qubits, which removes any leakage on the parity qubits, (3)~using a custom operation known as a \textit{LeakageISWAP} to remove data qubit leakage and move it to a parity qubit, and (4)~resetting the parity qubits yet again. The fidelity of this operation is rather high, as (1)~DQLR is not vulnerable to leakage transport, and (2)~it only requires a single two-qubit operation to remove leakage. However, as shown in Figure~\ref{fig:dqlrcircuit}(b), the DQLR protocol can introduce leakage on the data qubits if the first parity qubit reset fails (the parity qubit is initialized in $\ket{1}$ instead of $\ket{0}$), as then the data qubits may be excited to $\ket{2}$\footnote{This may occur as LeakageISWAP performs an $i$SWAP in the $\ket{11}, \ket{20}$ basis.}. Thus, much like SWAP-based LRCs, overusing the DQLR protocol is risky, as it may introduce leakage even when there was no leakage to begin with.

    \subsubsection{Results}
        We examine the applicability of \design\ to the DQLR protocol and assume that the LeakageISWAP gate has the same fidelity as a CX gate. We compare the baseline DQLR policy, which executes the leakage removal protocol every syndrome extraction round; \design\ and \designext, which schedule DQLR speculatively; and Optimal, which schedules DQLR whenever there is a data qubit leakage. 
        
\begin{figure}[!htb]
            \centering
            \includegraphics[width=\columnwidth]{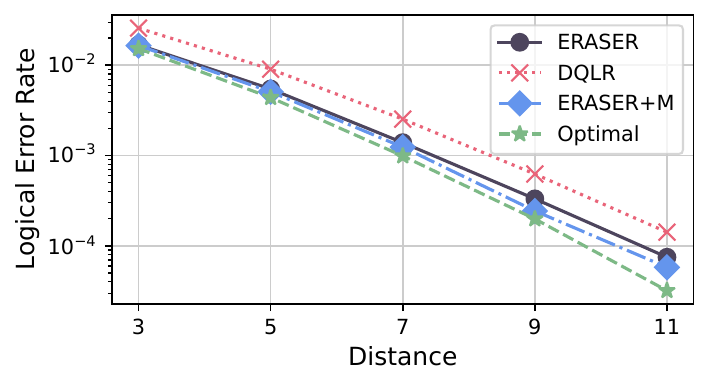}
            \caption{LER over $10$ QEC cycles at $p = 10^{-3}$ using DQLR instead of SWAPs.}
            \label{fig:dqlrler}
        \end{figure}

        Figure~\ref{fig:dqlrler} shows the LER for all four policies. We observe that \design\ improves upon the baseline DQLR protocol by $1.8\times$ on average and up to $1.9\times$, whereas \designext\ improves $2\times$ on average and up to $2.6\times$. We note that there is about a $4.4\times$ gap between the optimal scheduling of DQLR and the baseline DQLR protocol. These results demonstrate that custom approaches can benefit significantly from real-time scheduling.

\begin{figure}[!htb]
            \centering
            \includegraphics[width=\columnwidth]{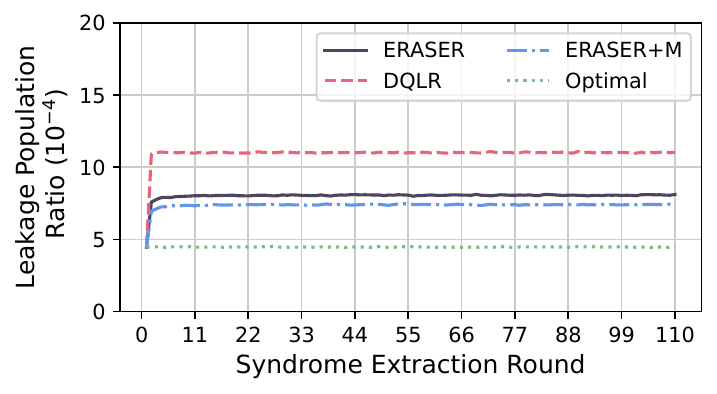}
            \caption{LPR over $110$ rounds for $d = 11$ at $p = 10^{-3}$ using DQLR instead of SWAPs.}
            \label{fig:dqlrlpr}
        \end{figure}

        We further examine the LPR of all four policies. Figure~\ref{fig:dqlrlpr} shows the LPR for all four policies for $d = 11$ over $110$ syndrome extraction rounds. Unlike SWAP-based LRCs, DQLR stabilizes the LPR rather quickly, as was reported in prior work~\cite{miao2022overcomingleakage}. However, as DQLR can cause leakage if the first reset fails, overusing DQLR can cause additional leakage. As \design\ and \designext\ judiciously schedule DQLR, they reduce the LPR by about $1.4\times$ and $1.5\times$ respectively. 
    
\section{Appendix: Artifact}

\subsection{Abstract}

The artifact contains the source code used to evaluate the designs proposed in this paper. We have listed how to reproduce the key results of our paper, namely those presented in Figures 5 and 6, which motivate the problem of inefficient LRC scheduling; Figures 14-16, which are our main results; and Table 2, which lists the utilization and timing results for our design (in RTL).

\subsection{Artifact check-list (meta-information)}

{\small
\begin{itemize}
  \item {\bf Algorithm: } ERASER, a leakage-detection algorithm. 
  \item {\bf Program: } \texttt{leakage}, \texttt{eraser\_rtl\_gen} 
  \item {\bf Compilation: } GCC 
  \item {\bf Hardware: } Tested on both Linux and MacOS 
  \item {\bf Execution: } Through command line 
  \item {\bf Metrics: } Logical Error Rate (LER) and Leakage Population Ratio (LPR)
  \item {\bf Output: } RTL, data files, and figures.
  \item {\bf Experiments: } Only three experiments (detailed later).
  \item {\bf How much disk space required (approximately)?: } At most 1 GB.
  \item {\bf How much time is needed to prepare workflow (approximately)?: } A minute or so of compilation.
  \item {\bf How much time is needed to complete experiments (approximately)?: } 48 hours
  \item {\bf Publicly available?: } Yes, on Zenodo
  \item {\bf Code licenses (if publicly available)?: } Apache
  \item {\bf Data licenses (if publicly available)?: } N/A
  \item {\bf Workflow framework used?: } N/A
  \item {\bf Archived (provide DOI)?: } 10.5281/zenodo.8224450 
\end{itemize}
}

\subsection{Description}

\subsubsection{How to access}

The artifact for this work is available on Zenodo at \url{https://doi.org/10.5281/zenodo.8224450}.

\subsubsection{Software dependencies}

The code is built using CMake v3.20.3, though slightly older versions should be fine and can be enabled by modifying \texttt{CMakeLists.txt}. The compiler used in our evaluations was g++-12 and g++-13, and we also used OpenMPI v4.x.x to parallelize the experiments on computing clusters. All other dependencies have been packaged with the code and are referenced through CMake.

The provided plotting script has been tested using Python v3.10.6, and the following packages are dependencies: matplotlib v3.6.1, numpy v1.23.4, scipy v1.9.2, though it should work fine with newer versions. 

For data involving RTL, we used Vivado 2023.1 to synthesize the design and obtain utilization and timing data, but older Vivado versions (i.e. 2022.x) should be sufficient.

\subsection{Installation}

We encourage using two build directories: \texttt{build} and \texttt{build\_RTL} to avoid any issues. \texttt{build} is for creating the data for Figures 5, 6, 14, 15, and 16, whereas \texttt{build\_RTL} is for generating the RTL (default distance $9$). The executables \texttt{leakage} and \texttt{eraser\_rtl\_gen} can be generated as follows:
\\
\\
{\small
\texttt{\$ cd build \\ \$ cmake .. -DCMAKE\_BUILD\_TYPE=Release \\ \$ make -j8 \\ \$ cd ../build\_RTL \\ \$ cmake .. -DRTL=On -DCMAKE\_BUILD\_TYPE=Release \\ \$ make -j8}}

\subsection{Experiment workflow}

\subsubsection{Main Paper Figures}

We explain how to generate the data for Figures 5, 6, 14, 15, and 16, which represent the main insights and results of the work. We have provided several bash scripts in the \texttt{leakage} folder: \texttt{figure\_5\_6.sh} and \texttt{figure\_14\_15\_16.sh} which generate the data for the corresponding figures. We recommend running \texttt{figure\_5\_6.sh} first as it can be done on any laptop within an hour. For \texttt{figure\_14\_15\_16.sh}, we recommend using a cluster with sufficient memory, as the larger distance codes require significant amounts of memory and may need many cores to complete in time. For reference, our evaluations for Figures 5 and 6 took five minutes on an ARM server using 64 cores using about 1GB per core. In contrast, our evaluations for Figures 14 through 16 took two days running on a cluster with $512$ cores, with about 8GB per core.

For \texttt{figure\_5\_6.sh}, there are only two parameters: \texttt{proc}, the number of processors (used by MPI), and \texttt{shots}, which is the number of trials to use in the experiment. The number of processors can be set to the user's preference. For \texttt{shots}, we used 100K in the paper, though 10K would be fine also.

For \texttt{figure\_14\_15\_16.sh}, there are three parameters: \texttt{p}, the physical error rate; \texttt{proc}, the number of processors; and \texttt{shots}, the number of trials in the experiment. In the paper, Figure 14 uses both $p = 10^{-3}$ and $p = 10^{-4}$, whereas Figure 15 and 16 both use $p = 10^{-3}$. We also note that the number of trials to obtain meaningful data \textit{increases} with code distance ($d$) and lower physical error rate. We found that 10M trials (\texttt{shots}) is sufficient for all experiments at $p = 10^{-3}$, whereas 100M trials will provide the data reported for $p = 10^{-4}$ in the paper. We note that $d = 9$ and $d = 11$ will be incomplete as they require more trials, likely 1B or beyond which is intractable to perform with our setup. 

In summary, to generate the data:
\\
\\
{\small
\texttt{\$ cd leakage \\ \$ ./figure\_5\_6.sh <PROC> 100000 \\ \$ ./figure\_14\_15\_16.sh 1e-3 <PROC> 10000000 \\ \$ ./figure\_14\_15\_16.sh 1e-4 <PROC> 100000000  }}
\\
\\
To plot the data, go to the \texttt{python} folder and call \texttt{plot.py}. This will generate PDFs for each figure in the \texttt{figures} folder.

\subsubsection{RTL Statistics}

To generate the RTL (which is in SystemVerilog), run:
\\
\\
{\small
\texttt{\$ cd build\_RTL \\ \$ ./eraser\_rtl\_gen <DISTANCE> > <RTL-FILE> }}
\\
\\
For example, \texttt{./eraser\_rtl\_gen 9 > eraser\_d9.sv} will write the RTL for a distance $9$ code to the \texttt{eraser\_d9.sv} file. After obtaining the RTL for distances $3$ to $11$, make a project in Vivado with the source file as the sole file. Then, add a constraint file (\texttt{.xdc}) to drive the \texttt{clk} signal for the RTL. Our file contained the single line\footnote{See \href{https://docs.xilinx.com/r/2021.1-English/ug895-vivado-system-level-design-entry/Adding-and-Creating-Constraint-Files}{\un{here}} for more details on adding constraint files in Vivado.}:
\\
\\
{\small
\texttt{create\_clock -name clk -period <PERIOD> -waveform \{0 <PERIOD/2>\} [get\_ports clk]}}
\\
\\
where \texttt{PERIOD} (which is in nanoseconds) can be set to any value based on the desired frequency (i.e. for a frequency of $250$MHz, set $\texttt{PERIOD} = 4$). Our designs generally have low critical path latencies, so high frequencies can be used. However, we believe a practical frequency would be $500$MHz (so $\texttt{PERIOD} = 2$). After defining the constraint file and adding it to the project, run \textit{Synthesis} on the project. Utilization and timing results can be obtained at this point by generating different reports for the project.

Our design does not require any IP blocks. We used the Kintex UltraScale+ FPGA to evaluate our design, and the specific part used for evaluations in the paper is \texttt{xcku3p-ffvd900-3-e}. 

To obtain RTL results for each distance, we recommend having one project that contains the RTLs for each distance. Then, obtaining results for each distance can be done by disabling the source files for other distances in Vivado. 

We note that for $d = 11$, our design requires more IO pins than available on the device. IO saturation is more an evaluation detail than a design detail. Rather, it is more practical that ERASER operates as a logic block in the larger fabric. Consequently, ERASER will likely not be interacting with external inputs. While the design will fail to synthesize for $d = 11$, timing and utilization data can still be retrieved from the design. If the user wants to avoid errors during synthesis, one can add \texttt{-mode out\_of\_context} to the project constraints file.

\subsection{Evaluation and expected results}

The results for LER and LPR should be about the same as the reported values in the paper, perhaps with slight deviations due to randomness. The results for RTL utilization and timing should be similar to that which was reported in Table 3, with some deviation due to randomness.

\subsection{Experiment customization}

To modify \design\ and \designext, see {\small\texttt{quarch/src/fleece.cpp}}. 
To modify the experiments, see {\small\texttt{leakage/src/experiments.cpp}}.

\subsection{Methodology}

Submission, reviewing and badging methodology:

\begin{itemize}
  \item \url{https://www.acm.org/publications/policies/artifact-review-and-badging-current}
  \item \url{http://cTuning.org/ae/submission-20201122.html}
  \item \url{http://cTuning.org/ae/reviewing-20201122.html}
\end{itemize}

\bibliographystyle{ACM-Reference-Format}
\balance
\bibliography{acmart}

\end{document}